\documentclass[structabstract]{aa}  

\usepackage{graphicx}
\usepackage{txfonts}
\usepackage{supertabular}
\usepackage{aalongtable}
\begin{document}
   \title{An HST/WFPC2 survey of bright young clusters in M31. IV.}

   \subtitle{Age and mass estimates.\fnmsep\thanks{Based on 
   observations made with the NASA/ESA Hubble Space Telescope, obtained at the
   Space Telescope Science Institute, which is operated by the Association 
   of Universities for Research in Astronomy, Inc., under NASA contract 
   NAS 5-26555. These observations are associated with program GO-10818 
   [P.I.: J.G. Cohen].}}

   \author{S. Perina\inst{1,2}, J.G. Cohen\inst{6}, P. Barmby\inst{3}, 
           M.A. Beasley\inst{4,5}, M. Bellazzini\inst{1}, 
	   J.P. Brodie\inst{4}, L. Federici\inst{1}, 
	   F. Fusi Pecci\inst{1}, S. Galleti\inst{1}, P.W. Hodge\inst{7}, 
	   J.P. Huchra\inst{8}, M. Kissler-Patig\inst{9}, 
	   T.H. Puzia\inst{10}\fnmsep\thanks{Plaskett Fellow.}
          \and
          J. Strader\inst{8}\fnmsep\thanks{Hubble Fellow.}
          }

\institute{INAF - Osservatorio Astronomico di Bologna, via Ranzani 1, 
           40127 Bologna, Italy\\
	   \and
Universit\`a di Bologna, Dipartimento di Astronomia, via Ranzani 1, 
           40127 Bologna, Italy\\
           \email{sibilla.perina2@unibo.it}\\
	   \and	   
Department of Physics and Astronomy, University of Western 
	      Ontario, London, ON, Canada N6A 3K7\\
	   \and	   
UCO/Lick Observatory, University of California, 
Santa Cruz, CA 95064, USA\\	  
	   \and	   
	      Instituto de Astrof�sica de Canarias, La Laguna 38200, 
	      Canary Islands, Spain\\ 
	   \and	   
	      Palomar Observatory, Mail Stop 105-24, California Institute of 
	      Technology, Pasadena, CA 91125\\
	      \email{jlc@astro.caltech.edu}\\
	   \and	   
	      Department of Astronomy, University of Washington, Seattle, 
	      WA 98195, USA\\
	   \and	   
	      Harvard-Smithsonian Center for Astrophysics, Cambridge, MA\\
	   \and	   
	      European Southern Observatory, Karl-Schwarzschild-Strasse 2, 
	      85748 Garching bei M\"unchen, Germany\\
	  \and    
	      Herzberg Institute of Astrophysics, 5071 West Saanich Road, 
	      Victoria, BC V9E 2E7, Canada\\
             }

   \date{Received, accepted }

 
  \abstract
   {}
   {We present the main results of an imaging survey of possible young massive 
   clusters (YMC)
   in M31 performed with the Wide Field and Planetary Camera 2 (WFPC2) on the  
   Hubble Space Telescope (HST), with the aim of estimating their age and their
   mass.  
   We obtained shallow (to B$\sim 25$) 
   photometry of individual stars in 19 clusters (of the 20 targets of the 
   survey).  We  present the images and color magnitude diagrams (CMDs) of all 
   of our targets.}
   {Point spread function fitting photometry of individual stars was
   obtained for all the WFPC2 images of the target clusters, and the 
   completeness of the final
   samples was estimated using extensive sets of artificial stars
   experiments. The reddening, age, and metallicity of the clusters were
   estimated by comparing the observed CMDs and luminosity
   functions (LFs) with theoretical models. Stellar masses were estimated by
   comparison with theoretical models in the 
   log(Age) $vs.$ absolute integrated magnitude plane, using ages estimated 
   from our CMDs and integrated J, H, K magnitudes from 2MASS-6X.}
   {Nineteen of the twenty surveyed candidates were confirmed to be real star
   clusters, while one turned out to be a bright star. Three of the clusters 
   were
   found not to be good YMC candidates from newly available integrated
   spectroscopy and were in fact found to be old from their CMD. Of the
   remaining sixteen clusters, fourteen have ages between 25 Myr and 
   280 Myr, two have older ages than 500 Myr (lower limits). By including 
   ten other YMC with HST photometry from the literature, we assembled a
   sample of 25 clusters younger than 1 Gyr, with mass ranging from $0.6\times
   10^4 M_{\sun}$ to $6\times 10^4 M_{\sun}$, with an average of 
   $\sim 3\times 10^4 M_{\sun}$. Our estimates of ages and masses well agree with 
   recent independent studies based on integrated spectra.}
   {The clusters considered  here are confirmed to have masses significantly 
   higher 
   than Galactic open clusters (OC) in the same age range. Our analysis indicates 
   that YMCs are relatively common in all the largest star-forming galaxies of 
   the Local Group, while the lack of known YMC older than 20 Myr in the Milky 
   Way may stem from selection effects.}
   

   \authorrunning{S. Perina et al.}
   \titlerunning{A HST/WFPC2 survey of bright young clusters in M31. II.}
   \maketitle
%

\section{Introduction}

Much of the star formation in the Milky Way is thought to have occurred within
star clusters  (Lada et al. \cite{lada91}, Carpenter et al.
\cite{carp}); therefore, understanding the formation and evolution of star
clusters is an important piece of the galaxy formation puzzle. Our
understanding of the star cluster systems of spiral galaxies largely comes
from studies of the Milky Way. Star clusters in our Galaxy have traditionally
been separated into two varieties, open and globular clusters  (OCs and GCs
hereafter).  OCs are conventionally regarded as young ($<10^{10}$ yr), 
low-mass  ($<10^4 M_{\sun}$), and metal-rich systems that reside in the
Galactic disk. 
In contrast, GCs are characterized as old, massive
systems. In the Milky Way, GCs can be broadly separated into two components: a
metal-rich disk/bulge  subpopulation, and a spatially extended, metal-poor halo
subsystem  (Kinman \cite{tom}, Zinn \cite{zinn}, see also Brodie \& Strader
\cite{brodie}, Harris \cite{h01}, for general reviews  of GCs).

However, the distinction between OCs and GCs has become increasingly blurred.
For example, some OCs are luminous and old enought to be confused with
GCs  (e.g., Phelps \& Schick \cite{phelps}). Similarly, some GCs are very 
low-luminosity systems (e.g., Koposov et al. \cite{kopos}), and at least one has
an age that is consistent  with the OC age distribution (Palomar 1, Sarajedini 
et al. \cite{sara}). 
Moreover, a third category of star cluster, ``young massive clusters'' (YMCs) 
are observed to exist in both merging  (e.g., Whitmore \&
Schweizer \cite{wischw}) and quiescent galaxies  (Larsen \& Richtler
\cite{lari}). Indeed, YMCs have been known to exist in the Large Magellanic
Cloud (LMC) for over half a century (Hodge \cite{ho61}).  These objects are
significantly more luminous than  OCs (M$_V\la-8$ up to M$_V\sim -15$), making
them promising candidate young GCs. Once thought  to be absent in the Milky
Way, recent observations suggest that their census may be quite incomplete, as
some prominent cases have been found  recently in the Galaxy as well 
(Clark et al. \cite{clark}, Figer \cite{figer}, Messineo et al. \cite{maria}).

Thus, a picture has emerged that, rather than being distinct groups, 
OCs, YMCs and GCs may represent regions within a continuum of cluster properties
dependent upon local galaxy conditions (Larsen \cite{lars}). The lifetime of a
star cluster is  dependent upon its mass and environment. Most low-mass star
clusters in disks are rapidly disrupted via interactions with giant molecular
clouds (Lamers \& Gieles \cite{lamers}, Gieles et al. \cite{gieles}).  
These disrupted  star clusters are
thought to be the origin of much of the present field star populations (Lada \&
Lada \cite{lala}). Surviving disk clusters may then be regarded as OCs or YMCs, 
depending upon their  mass. Star clusters in the halo may survive 
longer
since they are subjected to  the more gradual dynamical processes of two-body
relaxation and evaporation.  The clusters which survive for a Hubble time --
more likely to occur away from the  disk -- are termed GCs (see also
Krienke \& Hodge \cite{krie1}). 
To date, no known {\it thin} disk GCs have been identified in the Milky Way.

After the Milky Way, M31 is the prime target for expanding our knowledge of
cluster systems in spirals. 
However, our present state of knowledge about the M31
cluster system is far from complete.  Similar to the Milky Way, M31 appears to
have at least two GC  subpopulations, a metal-rich, spatially concentrated
subpopulation of GCs  and a more metal-poor, spatially extended GC
subpopulation (Huchra et al. \cite{huchra}, Barmby et al. \cite{barm00}). 
Also, again similar to
the Milky Way GCs, the metal-rich GCs in M31 rotate and  show "bulge-like"
kinematics (Perrett et al. \cite{perrett}). However, unlike the case in  the 
Milky Way, the metal-poor GCs also show significant rotation 
(Huchra et al. \cite{huchra}, Perrett et al. \cite{perrett}, 
Lee et al.~\cite{lee}).  
Using the Perrett et al. (\cite{perrett}) data, Morrison et al. (\cite{morri})
identified what  appeared to be a {\it thin} disk population of GCs, 
constituting some  27\% of the Perrett et al. (\cite{perrett}) sample. 
Subsequently, it has been shown that at least a subset of these objects  are 
in fact young ($\leq$ 1 Gyr), metal-rich star clusters rather than 
old ``classical''  GCs (Beasley et al. \cite{beas}, 
Burstein et al. \cite{burst}, Fusi Pecci et al.~\cite{ffp}, 
Puzia et al. \cite{puzia}, Caldwell et al. \cite{C09}).

Fusi Pecci et al. (\cite{ffp}, hereafter F05) presented a comprehensive 
study of bright young disk clusters in M31, selected from the Revised
Bologna Catalog\footnote{\tt www.bo.astro.it/M31} (RBC, Galleti et
al.~\cite{rbc}) by color [$(B-V)_0\le 0.45$] or by the strength of the
$H\beta$ line in their spectra ($H\beta\ge 3.5\AA$). While these
clusters have been noted since Vetesnik (\cite{vetes}) and have been
studied by various
authors, a systematic study was lacking. 
F05 found that these clusters, that they termed -- to add to the growing 
menagerie of star cluster species -- ``blue luminous compact clusters'' 
(BLCCs), are fairly numerous in M31 (15\% of the whole GC sample), they have 
positions and kinematics typical of {\em thin disk} objects, and
their colors and spectra strongly suggest that they have ages
(significantly) less than 2 Gyr. 

Since they are quite bright ($-6.5\la M_V\la -10.0$) and -- at least in some
cases -- morphologically similar to old GCs (see Williams \& Hodge
\cite{will}, hereafter WH01), BLCCs could be regarded as YMCs, that is 
to say, candidate  young GCs (see De Grijs \cite{degris2}, for 
a recent review). 
In particular, F05 concluded
that if most of the BLCCs have an age $\ga 50-100$ Myr they are likely
brighter than Galactic open clusters (OC) of similar ages, thus they
should belong to a class of objects that is not present, in large numbers, in 
our own Galaxy. 
Unfortunately, the accuracy in the age estimates obtained from the 
integrated properties of the clusters is not sufficient to determine their 
actual nature on an individual basis, i.e., to compare their total luminosity 
with the luminosity distribution of OCs of similar age (see 
Bellazzini et al. \cite{cefa}, hereafter B08, and references therein). 

In addition to the question of the masses and ages of these BLCCs, it has
become clear that the BLCC photometric and spectroscopic samples in M31 may
suffer from significant contamination.  Cohen, Matthews \& Cameron (\cite{coh}, 
hereafter C06) presented NIRC2@KeckII Laser Guide Star Adaptive Optics (LGSAO) 
images of six candidate BLCCs. Their $K\arcmin$ very-high spatial resolution 
images revealed that in the fields of four candidates there was no
apparent cluster. This led C06 to the conclusion that some/many of the claimed
BLCC may in fact be just {\em asterisms}, i.e. chance groupings of stars in the
dense disk of M31. The use of the near infrared $K\arcmin$ band
(required by the LGSAO technique) may be  largely insensitive  to very young
clusters that are dominated by relatively few hot stars, which emit most of 
their
light in the blue region of the spectrum. Hence, the imaging by C06 may be
inappropriate to detect such young clusters (see, for example, the detailed
discussion by Caldwell et al. \cite{C09}).  
In any case, the study by C06 suggests that the true number
of massive young clusters of M31 may have been overestimated.

   \begin{figure*}
   \centering
   \includegraphics[width=18cm]{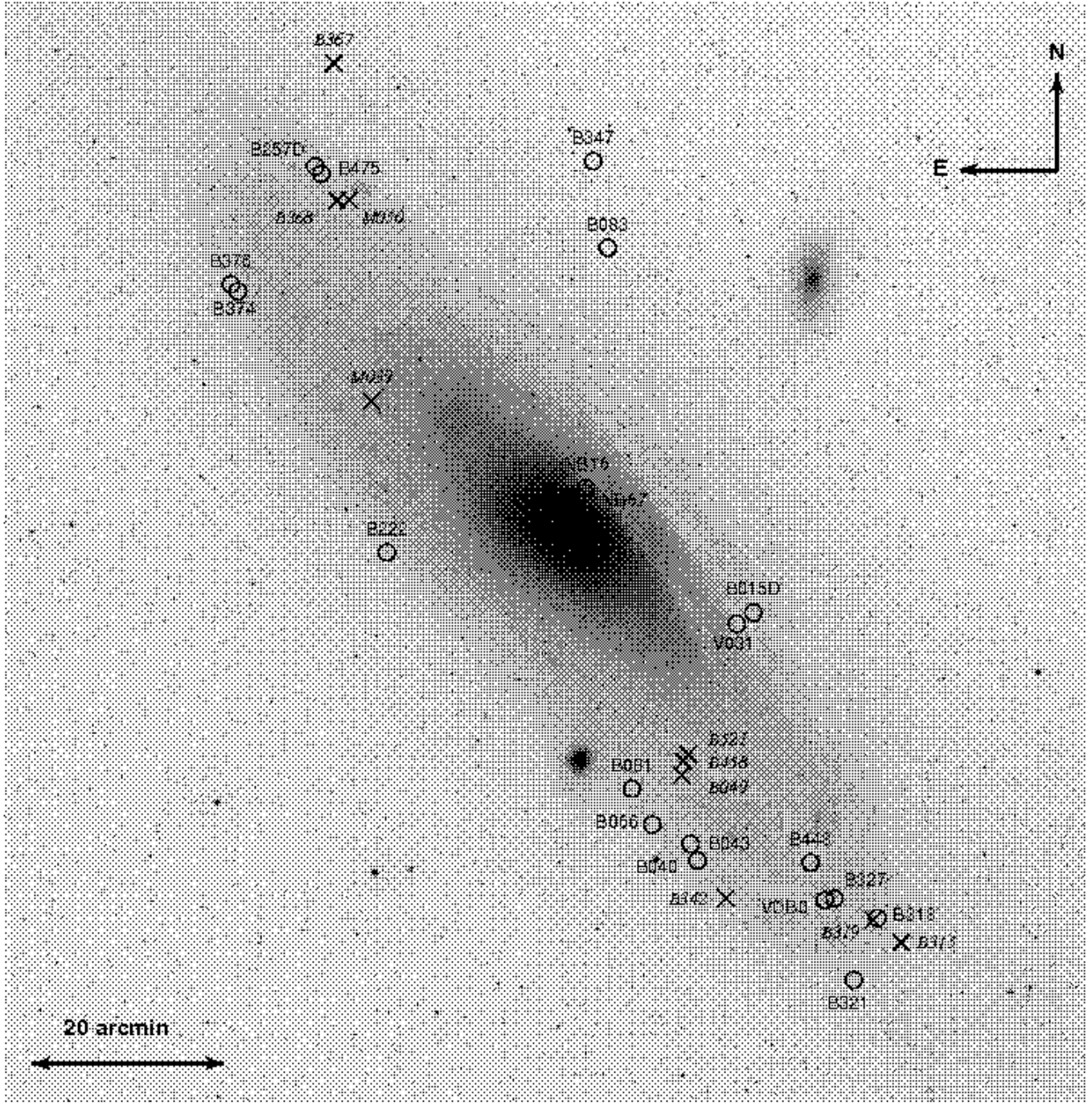}
      \caption{Location of the 20 targets of our survey (empty circles) 
      projected against the body of M31. The $\times$ symbols indicate the 
      position of the additional ten Young Clusters we included 
      in Sect.~\ref{massageir}.}
         \label{cpos}
   \end{figure*}

Therefore, in order to ascertain the real nature of these BLCCs we have
performed  a survey with the Hubble Space Telescope (HST) to image 20 BLCCs in the disk of M31 (program 
GO-10818, P.I.: J. Cohen).  The key aims of the survey are:

\begin{enumerate}

\item to check if the imaged targets are real clusters or asterisms, and
to determine the fraction of contamination of BLCCs by asterisms, 

\item to obtain an estimate of the age of each  cluster in order to verify
whether it is brighter than Galactic OCs of similar age. Ultimately the
survey aims to provide firm conclusions on the existence of a significant 
population of BLCCs (YMCs) in M31, in addition to OCs (see
Krienke \& Hodge \cite{krie1,krie2}, and references therein) and GCs. 

\end{enumerate}

   \begin{figure*}
   \centering
   \includegraphics[width=18cm]{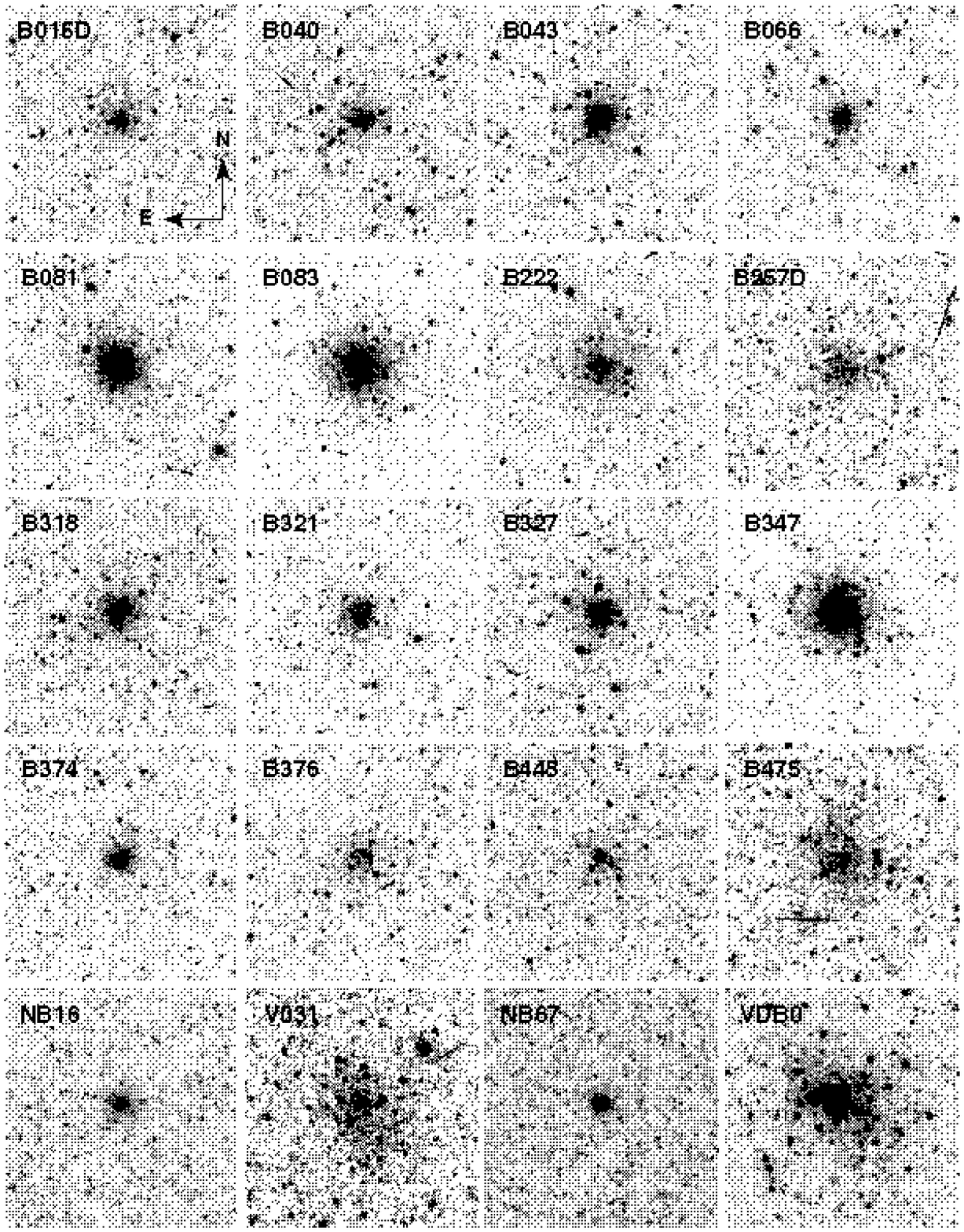}
      \caption{F450W images of the 20 primary targets. 
      Each image covers the central 10$^{\arcsec}\times$~10$^{\arcsec}$
      on the PC field 
      (10$^{\arcsec}$ = 38 pc at the assumed M31 distance modulus of 24.47). 
      North is up and East to the left.}
         \label{cima}
   \end{figure*}
%


%
 
\begin{table*} 
\caption{Positional, photometric and spectroscopic parameters for the 
surveyed clusters.}

\label{table:1}      
\centering 
\begin{tabular}{l c c c c c c c c c c c c}      

\hline\hline          
Name & X$^a$ & Y$^a$ & R &  B &  V & (B-V)$_{0}$$^{F05}$ & (B-V)$_{0}$$^{(t.w.)}$ & H$_{\beta}$$^{F05}$ & H$_{\beta}$$^{G09}$ & ff$^b$ \\
&(arcmin)&(arcmin)&(arcmin)&&&&&(\AA)&(\AA)&\\

\hline 
\\
 B015D-D041 &  -19.27 &   9.22 &  21.36  & 19.11$\pm$0.02 & 18.36$\pm$0.03 & $\dots$ &   0.15 & 7.32	   & $\dots$  & 1 \\
 B040-G102  &  -35.40 & -11.92 &  37.35  & 17.54$\pm$0.03 & 17.20$\pm$0.04 & 0.18    &   0.11 & 7.41	     & 7.58$\pm$ 0.30	 & 1 \\
 B043-G106  &  -33.62 & -11.37 &  35.49  & 17.04$\pm$0.03 & 16.77$\pm$0.04 & 0.17    &   0.04 & 5.53	     & 5.70$\pm$ 0.30	 & 1 \\
 B066-G128  &  -29.55 & -13.17 &  32.35  & 17.56$\pm$0.03 & 17.35$\pm$0.04 & 0.25    &  -0.02 & 4.67	     & 4.84$\pm$ 0.30	 & 1 \\
 B081-G142  &  -25.26 & -12.36 &  28.12  & 17.36$\pm$0.02 & 16.86$\pm$0.03 & 0.43    &    0.20 & 7.98	     & 8.15$\pm$ 0.30	 & 1 \\
 B257D-D073 &   45.98 &   4.02 &  46.16  & 18.41$\pm$0.02 & 18.00$\pm$0.04 & $\dots$ &   0.01 & 5.49	   & 5.66$\pm$ 0.30   & 1 \\
 B318-G042  &  -52.14 &  -1.32 &  52.16  & 17.02$\pm$0.03 & 16.82$\pm$0.03 & 0.06    &   0.03 & $\dots$	& 5.49$\pm$ 0.12    & 1 \\
 B321-G046  &  -55.50 &  -7.41 &  55.99  & 17.82$\pm$0.02 & 17.51$\pm$0.03 & 0.11    &   0.06 & 6.29	     & 6.85$\pm$ 0.32	 & 1 \\
 B327-G053  &  -47.67 &  -3.45 &  47.79  & 16.75$\pm$0.03 & 16.58$\pm$0.03 & 0.21    &  -0.03 & 4.09	     & 3.78$\pm$ 0.14	 & 1 \\
 B376-G309  &   42.16 & -10.67 &  43.49  & 18.35$\pm$0.02 & 17.97$\pm$0.04 & 0.34    &   0.08 & $\dots$	& 6.40$\pm$ 0.06    & 1 \\
 B448-D035  &  -43.16 &  -2.97 &  43.26  & 18.01$\pm$0.03 & 17.46$\pm$0.04 & 0.50    &    0.20& 6.70	     & 6.87$\pm$ 0.30	 & 1 \\
 B475-V128  &   45.00 &   4.06 &  45.18  & 17.55$\pm$0.03 & 17.09$\pm$0.04 & 0.20    &   0.11 & 5.96	     & 6.13$\pm$ 0.30	 & 1 \\
 V031       &  -19.03 &   7.17 &  20.34  & 18.16$\pm$0.03 & 17.62$\pm$0.04 & 0.57    &   0.19 & 5.84	     & 6.01$\pm$ 0.30	 & 1 \\
 B083-G146  &   19.83 &  22.08 &  29.68  & 17.85$^{d}$    & 17.09$^{d}$    & 0.65    &   0.56 & 3.75     & 1.75$\pm$ 0.42    & 1 \\
 B222-G277  &   10.22 & -16.16 &  19.12  & 18.00$\pm$0.02 & 17.24$\pm$0.03 & 0.57	  &   0.56 & 8.47	 & 4.46$\pm$ 0.31    & 1 \\
 B347-G154  &   27.74 &  26.74 &  38.53  & 17.23$^{d}$    & 16.50$^{d}$    & 0.62    &   0.67 & $\dots$  & 2.87$\pm$ 0.17    & 2 \\
 B374-G306  &   41.13 & -10.55 &  42.46  & 18.69$\pm$0.03 & 18.23$\pm$0.04 & 0.33	  &   0.16 & 4.07	   & 4.24$\pm$ 0.30    & 1 \\
 NB16       &    1.96 &   4.19 &   4.63  & 18.83$\pm$0.04 & 17.59$\pm$0.10 & 0.55	  &   0.99 & $\dots$	& 3.34$\pm$ 0.08    & 2 \\
 VDB0       &  -47.16 &  -4.33 &  47.36  & 14.94$\pm$0.09$^{c}$ & 14.67$\pm$0.05$^{c}$ & 0.12	  &   0.07 & 4.30   & 4.50$\pm$ 0.07   & 1 \\  
 NB67-AU13  &    1.68 &   3.73 &   4.09  & 16.48$\pm$0.02 & 15.92$\pm$0.03 & 0.37	  &   0.36 & $\dots$	 & $\dots$   & 1 \\
\hline                                                                                                           
\hline                                                                                                           
\end{tabular}                                                                                                    
\begin{list}{}{}  
\item B and V magnitudes are from new aperture photometry performed on 
the CCD images of Massey et al.~(\cite{massey}), except for B083 and B347 that 
are not included in the area covered by that survey.                                                                                               
\item[$^{\mathrm{a}}$]  X and Y are projected coordinates in the direction along                                 
(increasing Eastward) and perpendicular to the major axis of M31, in arcmin.
\item[$^{\mathrm{b}}$] ff is a flag indicating if the target has been selected from
Table 1 or Table 2 of F05.
\item[$^{\mathrm{c}}$] From Pap-I. 
\item[$^{\mathrm{d}}$] From the RBC.
\item[$^{\mathrm{(t.w.)}}$] from this work: B and V from this table and 
E(B-V) as estimated in Sect.~3 from isochrone fitting.
\item[$^{\mathrm{F05}}$] from Fusi Pecci et al. (2005): (B-V)$_{0}$ are 
calculated assuming a single value of E(B-V)=0.11 for all the clusters.
\item[$^{\mathrm{G09}}$] from Galleti et al. (2009).
\end{list} 
\end{table*}
%

In Perina et al. (\cite{P09a}, hereafter Pap-I) we have described in detail the
observational material coming from our survey, and the data reduction and
methods of analysis that we homogeneously adopt for the whole survey. We did
that by taking the brightest of our surveyed clusters (VdB0) as an example. In
this contribution we apply the same process to the whole sample, 
obtaining metallicity, reddening and age estimates for all the targets of our
survey. We incremented our final sample of candidate M31 YMC by including in the
final analysis ten further  clusters having age estimates  available from the 
literature that are fully homogeneous with our own ones. In two companion 
papers, Hodge et
al.~(\cite{lessclu}, Pap-II, hereafter) identified and studied clusters of 
lower mass (with respect to
those studied here) that were serendipitously imaged in our 
survey, while Barmby et al.~(\cite{bar09}, Pap-III, hereafter) studied 
the structure of the clusters
that are the main targets of the survey.

The paper is organized as follows. 
The sample is described in detail in Sect.~2, where we also summarize the data
reduction procedure. In Sect.~3 we present the individual color magnitude diagrams (CMDs) and luminosity
functions (LFs), we estimate ages, metallicities and reddening of each cluster. 
In Sect.~4 we derive the mass estimates for the clusters of our extended sample
(including data from the literature), we compare our clusters with open and
globular clusters of the Milky Way and we compare our estimates with those from
the recent and extensive analysis of young M31 clusters by Caldwell et al. 
(\cite{C09}, hereafter C09), that are based on integrated spectra. 
In Sect.~5 our main results are briefly summarized and discussed. 
Finally, in Appendix A we report on M31 clusters or candidate clusters listed 
in the RBC that have been serendipitously imaged within our survey, and, in 
Appendix B, we report on the nature of candidate BLCC=YMC M31 clusters that 
have an HST image in the archive, independent of this survey.


\section{Description of the sample}
\label{sample}

Table~\ref{table:1} lists the target clusters of our survey and reports
some positional and spectro-photometric parameters that were relevant for their
selection. New homogeneous large-aperture ($r_{ap}\sim 5\arcsec-10\arcsec$, depending on the curve of growth of each cluster) integrated B,V photometry for all the targets has been obtained from the publicly available CCD images by Massey et al.~(\cite{massey}), and calibrated using the published photometry from the same authors, as done in Pap-I for VdB-0 (see Pap-I for further details). 

Fig.~\ref{cpos} shows that the vast majority of the targets are projected
onto the so-called {\em 10 kpc ring} (see Hodge \cite{hodge}, Barmby et al.
\cite{spitzer}, C09 and references therein), a site of ongoing star formation in
the thin disk of M31. The only exceptions are B347 and B083, that are
significantly farther from the center of the galaxy, and NB16 that is projected
onto the outer regions of the M31 bulge. 
We will see below that these three clusters do
not fulfill the selection criteria by F05 for {\em bona fide} candidate YMCs and, 
in fact, they are likely old (see Sect.~\ref{terzi}).

Eighteen of the twenty targets were drawn from Tab.~1 of F05, i.e. they were
confirmed clusters\footnote{RBC class f=1, meaning that they have been classified as bona-fide M31 clusters by some author, based on their spectra and/or high resolution images.} that were classified as genuine BLCC = YMC by  
these authors as they
had $H_{\beta}\ge 3.5\AA$ or, when lacking a measure of $H_{\beta}$, 
$(B-V)_0\le 0.45$. After a careful inspection of the HST archive, we 
excluded from the selection any cluster from Tab.~1 of F05 that had 
already been imaged with HST (serendipitously, in most cases, see Appendix B), 
and we chose the brightest 18 among the remaining ones.
F05 assumed E(B-V)$=0.11$ for all the considered sample,
in Sect.~3 we will show that the typical reddening of these clusters is significantly
higher than this, in most cases E(B-V)$\ge 0.20$, in good agreement with the
estimates by C09 (see Fig.~\ref{confr_ebv}).  Hence, in general, the
$(B-V)_0$ colors derived here are bluer than those adopted by F05. Galleti et
al. (2009, G09 hereafter) presented new estimates of the $H_{\beta}$  index
(with respect to those reported by F05), taken either from their own 
observations or from the recent literature. In Table~\ref{table:1} we report
both the $(B-V)_0$ and $H_{\beta}$ values from F05 (that were used for the
selection of the sample) and those derived here and in G09, when available\footnote{Note that the scales of the $H_{\beta}$ index adopted by F05 and G09 are slightly different. The $H_{\beta}\ge 3.5\AA$ threshold by F05 translated into
$H_{\beta}\ge 3.7\AA$ in the scale by G09 (see the latter paper for discussion and details).}.
In one case (B083) the new value of $H_{\beta}$ is much lower than that reported
by F05 (1.75$\AA$ instead of 3.75$\AA$) and than the selection limit. 
Moreover, even with the new E(B-V) estimate derived here, $(B-V)_0=0.551$,
significantly redder that the limit adopted for the selection.
For these reasons B083 can no longer be considered as a 
candidate YMC, as it does not fulfill the selection criteria when the newly
available data are considered. The analysis of the CMD (in Sect.~3) will confirm 
that the cluster is in fact much older than genuine YMC, and possibly as old as
classical GCs.

The remaining two targets (NB16 and B347) were selected form Tab.~2 of F05,
including clusters not fulfilling their selection criteria for YMC but
classified as young (or possibly young) by some author in the past.  In both
cases  $H_{\beta}$ were lacking at the time, and the new values reported by G09
are  significantly below the selection threshold for a YMC. B347 is also much redder than 
$(B-V)_0= 0.45$. On the other hand, we find $(B-V)_0= 0.399$ for NB16. In this
case the criterion based on $H_{\beta}$ must prevail over that based on
de-reddened color as the former is reddening-independent, while relatively low
photometric and/or reddening errors can shift the color of this cluster above or below the
selection threshold. In conclusion, the newly available data indicates that 
both NB16 and B347  are  not good YMC candidates, as will be confirmed by their
CMDs (see Fig.~\ref{ageI}).  Hence, just re-considering the original selection in the light of
new estimates of integrated properties, our sample of {\em bona fide} YMC
candidates is reduced to 17 objects, including VdB0 which was studied in detail
in Pap~I.  

Postage stamp images of all the targets, from our HST data, are presented in
Fig.~\ref{cima} (see Sect.~\ref{obs}). 
Inspection of the images reveal that all our targets are
actually {\em genuine} clusters, with the only exception of NB67 that is a
bright star projected into a dense background of M31 (disc) stars (see also Pap-III, for the light profiles of the clusters).
For obvious reasons NB67 will be not considered further in the following analysis.
A first conclusion that can be drawn just from this preliminary analysis is that
the incidence of spurious objects in our sample is of 1/17$\simeq$ 6\%, much
lower than hypothesized by C06. If we consider the set of 
36 objects listed by F05 in their Tab.~1 for which  HST images were 
available in the archive we obtain the same result (see Appendix~\ref{appB}, for discussion
and further details).
Moreover, none of the considered clusters is in
fact an {\em asterism} (including those considered in 
Appendix~\ref{appB})\footnote{Bright stars are well-known classical
contaminants in lists of candidate M31 clusters of any kind, see Galleti et al.
\cite{svr}.}. Finally, if we extend our analysis to all the objects classified as YMC by F05 that have been ever imaged with HST we find the same very low degree of contamination (see Appendix~\ref{appB}).
Hence we are dealing with a significant class of real
stellar systems. A second conclusion is that while some of the considered
cluster appear quite extended and sparse (like, for example, B257D, B475, and 
V031), there are also rather compact globular-like clusters (like, B043, 
B081, and B327, as noted earlier B347 is likely old). 

   \begin{figure}
   \centering
   \includegraphics[width=9cm]{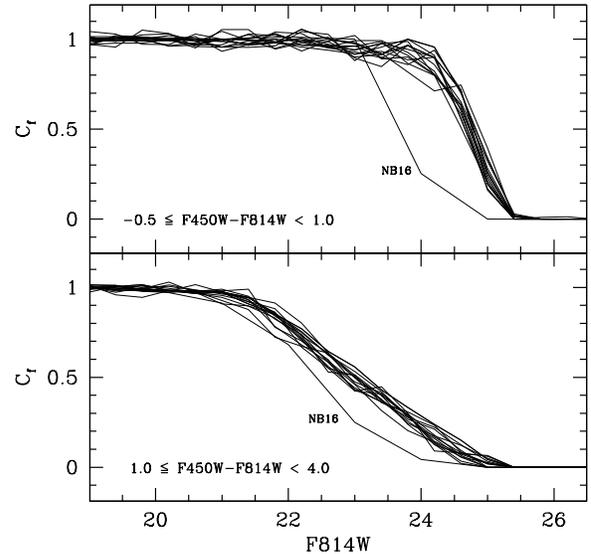}
      \caption{
      Completeness ($C_f$) of the samples as a function of F814W magnitude,
      obtained from artificial stars experiments, for
      all the clusters of our survey (listed in Tab.~{table:1}) and for two
      different color ranges. The upper panel is for a color range enclosing the
      MS of young clusters, the lower panel is for a color range
      enclosing the red giant stars. The $C_f(F814W)$ function of each cluster
      (for each color range) is computed considering only artificial stars
      enclosed in the radial range that is used to select the
      sample dominated by cluster stars that will be studied in the following 
      (typically $r\le 5\arcsec$, 
      see Sect.~\ref{profsec} and Sect.~\ref{secabe}). Note that all the
      $C_f(F814W)$ functions are very similar, except for the case of the
      exceedingly compact (and crowded) cluster NB16, labeled in both panels.  
            }
         \label{compl}
   \end{figure}
%

\subsection{Observations, data reduction and assumptions.}
\label{obs}

The characteristics of the survey data and the whole process of data reduction
and data analysis that has been applied in this study is described in
detail in Pap-I. In these section we briefly summarize the key characteristics
of the dataset and of the process, for the convenience of the reader.

Two $t_{exp}=400~s$ images per filter (F450W and F814W) were acquired for each
cluster with the Wide Field and Planetary Camera (WFPC2) on board of HST,
keeping the target at the center of the PC field. Unlike the case of VdB0,
treated in Pap-I, the clusters studied here have limiting radii significantly
smaller than the size of the PC camera ($\simeq 39\arcsec \times 39\arcsec$, see Pap-III), 
therefore both the cluster population
and the surrounding field can be studied using the PC images alone (see
Sect.~\ref{profsec}) without relying on the WF cameras. The
analysis of the field population in the portions of the M31 disk sampled by our
WF images will be the subject of another contribution (Perina et al., in
preparation). 

Photometry of the individual stars has been obtained with HSTPHOT (Dolphin
\cite{hstphot}), a Point Spread Function fitting package specifically 
developed for WFPC2 data. The reduction process includes cleaning of
cosmic-ray hits and bad pixels, correction for Charge Transfer Efficiency (CTE,
Dolphin \cite{dolcal}), and absolute photometric calibration in the VEGAMAG
system (Holtzman et al. \cite{holtz}, Dolphin \cite{dolcal}). The images were
searched for sources having peak intensities at $3\sigma$ above the 
background. The output catalogs were cleaned of spurious and/or badly measured
sources by selecting stars with HSTPHOT global quality flag=1, {\em crowding}
parameter $<0.3$, $\chi^2<2.0$ and $|sharp| <0.5$. 
The final catalogs containing position and F450W, F814W photometry of the PC
fields will be made publicly available through a dedicated 
WEB page\footnote{\tt www.bo.astro.it/M31/YMC}.

We estimated the completeness of our samples as a function of magnitude, color
and position on the field by means of extensive artificial stars experiments 
(more than $10^5$ artificial stars were simulated, per field of view, i.e.
more than $4\times 10^5$ per cluster), as described in detail in Pap-I.
Fig.~\ref{compl} show the completeness factor ($C_f$) as a function of magnitude for all
the clusters, for two different color ranges (one covering the clusters' main
sequence (MS) and one covering the Red (Super) Giant branches). The reported $C_f$
curves refers to the circles enclosing most of the cluster population that
are defined in Sect.~\ref{profsec}, hence they are fully relevant for the
following analysis. Note that the completeness conditions are very similar for
all the clusters (including VdB0, presented in Pap-I), except NB16. This cluster
is so compact that the considered region is much more crowded than all the other
cases, thus the completeness is significantly worse. 
The typical photometric uncertainties as derived from the artificial stars 
experiments are $\la \pm 0.02$ for $F450W\simeq F814W\le 21$, 
$\la \pm 0.05$ for $F450W\simeq F814W\le 22.5$, and $\la \pm 0.2$ for $F450W\simeq
F814W\le 24.0$ (see Pap-I, for details).

In the following we will always assume $(m-M)_0=24.47$, from McConnachie et al.
(\cite{dista}), corresponding to $D=783$ kpc. At this distance $1\arcsec$
corresponds to 3.8 pc, $1\arcmin$ to 228 pc. We adopt $A_{F450W}=4.015E(B-V)$
and $A_{F814W}=1.948E(B-V)$, from Schlegel et al. (\cite{dirbe}).
We will use theoretical isochrones and LFs in the HST/WFPC2
VEGAMAG system from the set by Girardi et al. (2002, hereafter G02), considering
only models in the range of metallicity $\frac{2}{5} Z_{\sun}\la Z\la
2Z_{\sun}$, that seem appropriate for young disk clusters.
Details and discussion regarding the choices outlined above can be found in Pap-I.

   \begin{figure}
   \centering
   \includegraphics[width=9cm]{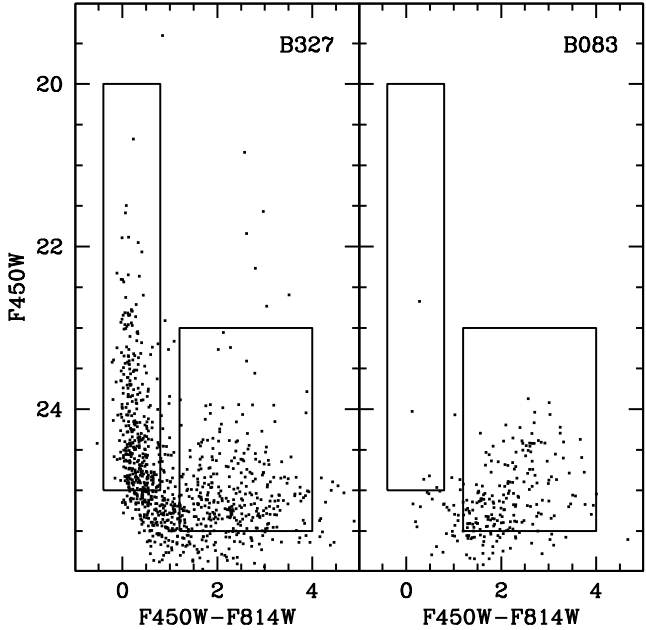}
      \caption{
      Selection boxes used for the stellar surface density 
      profiles shown in Fig.~\ref{prof1} and \ref{prof2}, 
      are superimposed on the CMD of two of the surveyed
      clusters taken as examples: a young cluster with a prominent MS (left
      panel) and an older cluster displaying just the tip of the RGB (right
      panel). The blue box at $F450W-F814W\sim 0.5$ selects bright MS stars 
      (young population), the faint redder box ($F450W-F814W> 1.0$) selects
      red giant stars (old population). In a few cases, the boxes have been
      slightly shifted in color to best match the MS and RGB features of
      a cluster
      with higher reddening. 
      }
         \label{boxes}
   \end{figure}
%

\subsection{Radial selection and first classification}
\label{profsec}

\begin{figure*}
   \centering
   \includegraphics[width=\textwidth]{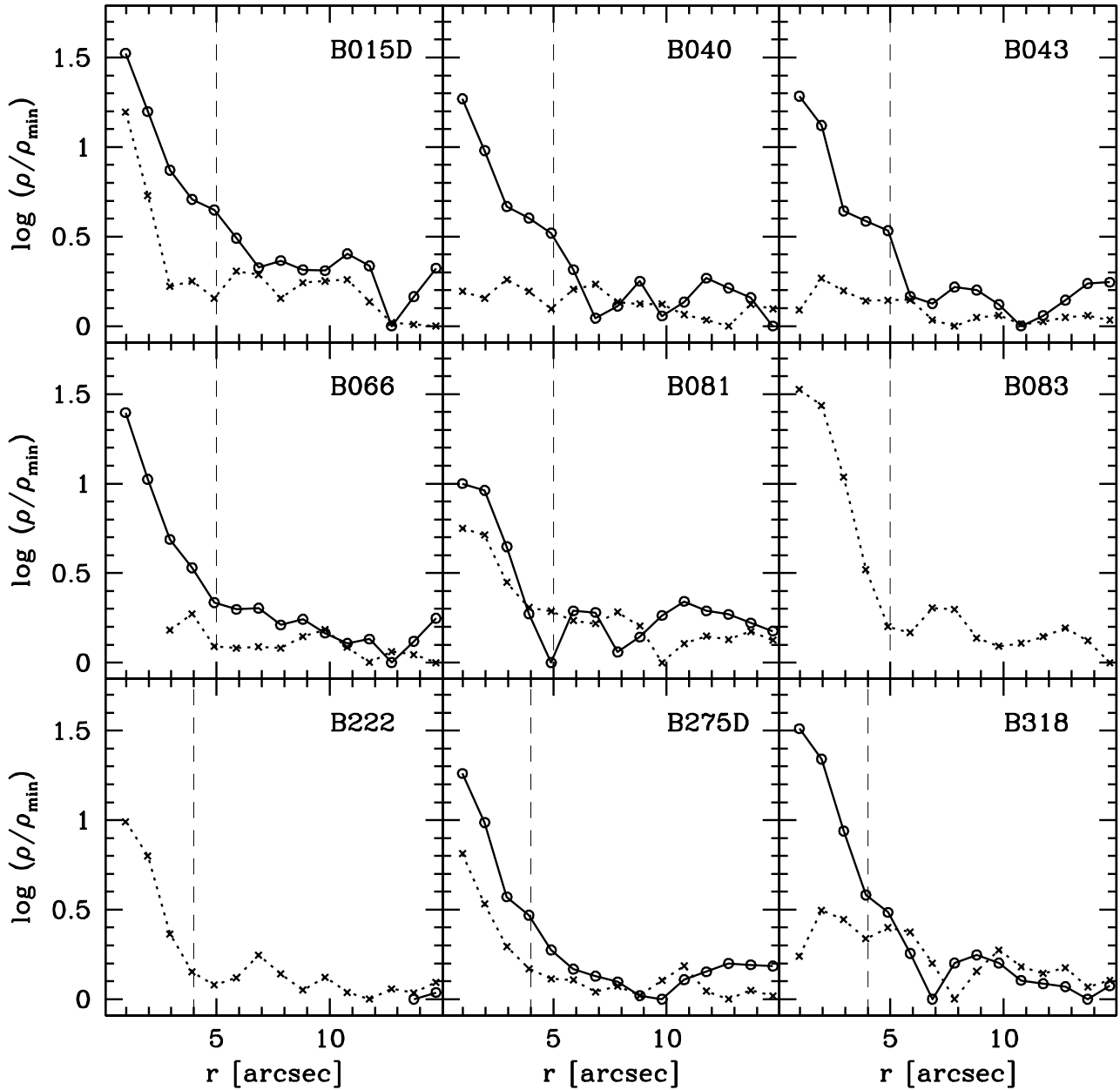}
      \caption{Stellar surface density profiles of the young 
      (open circles connected by a continuous line)
      and old (crosses connected by a dashed line) populations (as defined
      by the selection boxes illustrated in Fig.~\ref{boxes})  
      for nine of the surveyed clusters. }
         \label{prof1}
   \end{figure*}

\begin{figure*}
   \centering
   \includegraphics[width=\textwidth]{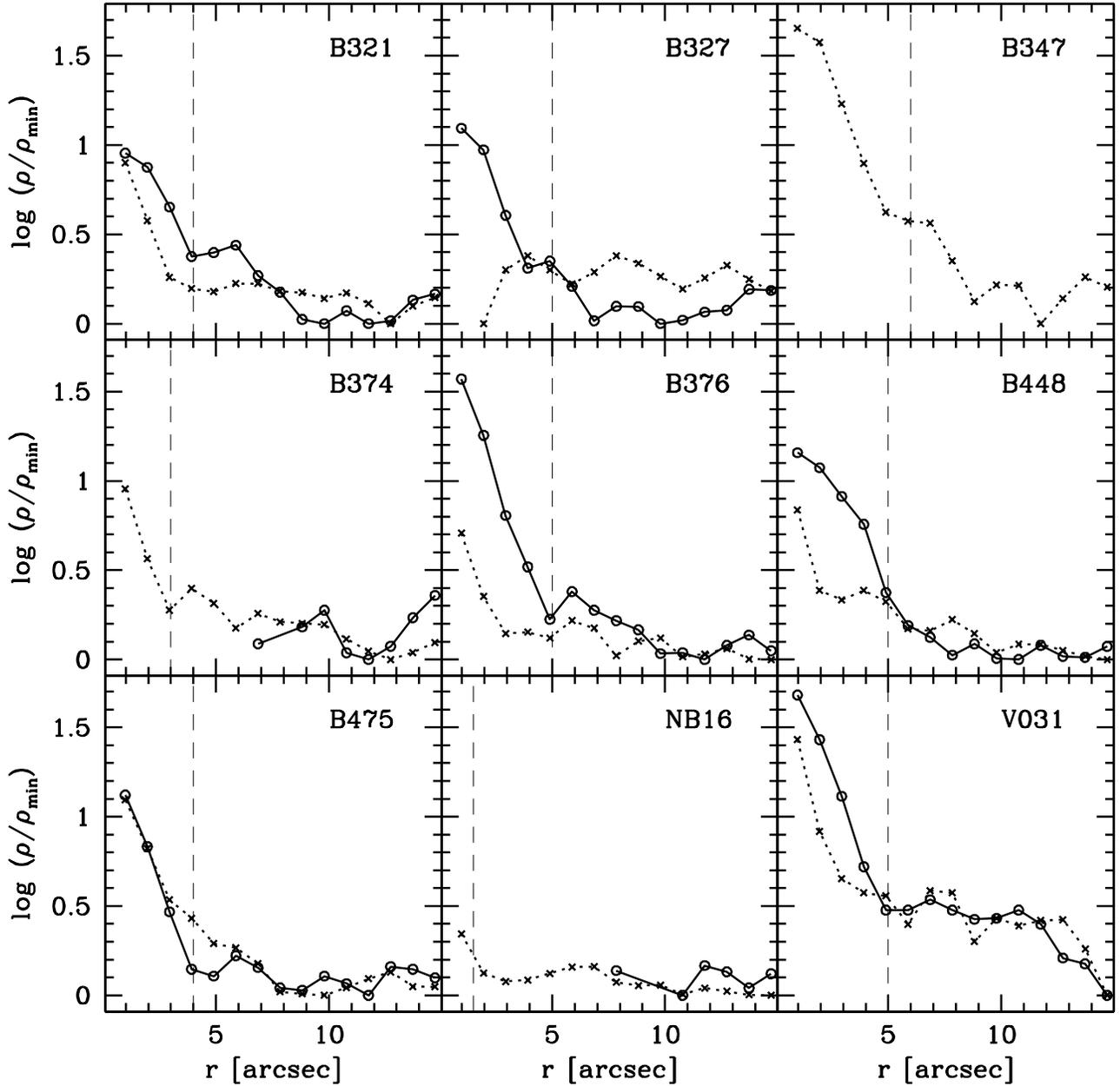}
      \caption{Same as Fig.~\ref{prof1} for the remaining nine 
      surveyed clusters.}
         \label{prof2}
   \end{figure*}

Before proceeding with the analysis of the CMDs of the clusters, we
need to select - for each cluster - a sub-sample of the PC field that is as representative as
possible of the cluster population, possibly minimizing the contamination by the
surrounding M31 field. Following Pap-I we adopt a radial selection, retaining in
the final {\em cluster sample} the stars lying within a certain distance from the
cluster center. To determine the selection radius to be adopted for each
individual cluster we proceeded as follows:

\begin{itemize}

\item We defined two broad selection boxes on the CMD, one enclosing the bright
MS typical of young clusters (Blue Box) and one enclosing a 
redder region
that should be dominated by old stars at the tip of the red giant branch (RGB) 
but can enclose also intermediate-age asymptotic giant branch (AGB) and some 
red super giant (RSG) stars, as illustrated in Fig.~\ref{boxes} (Red Box).

\item We derived surface-density radial profiles by counting stars selected in 
the two boxes on concentric annuli. To obtain smoother profiles with the relatively low number of stars available we adopted overlapping annuli of width 
$1.8\arcsec$, with a radial step of $0.9\arcsec$ between subsequent annuli.
The profiles from main sequence (MS)
stars and from red stars (shown in Fig~\ref{prof1} and
\ref{prof2}) are normalized to the minimum surface-density
encountered in the raster of radial annuli, that should be considered as roughly
representative of the surrounding field. For example, the profiles of B066, 
in the middle left panel of  Fig~\ref{prof1}, shows that at
the center of this cluster the surface density of bright MS stars is $\ga 20$
times higher than in the surrounding field, while there is no overdensity of red
stars correlated to the cluster.

\item Based on the scale of the detected overdensity we fixed the selection
radius of each cluster (marked in the plots as a vertical dashed line), with
the aim of isolating a circle that should be dominated by cluster stars. The
typical selection radius is $r\sim 5\arcsec$.

\end{itemize}

%

%

 
   \begin{figure*}
   \centering
   \includegraphics[width=\textwidth]{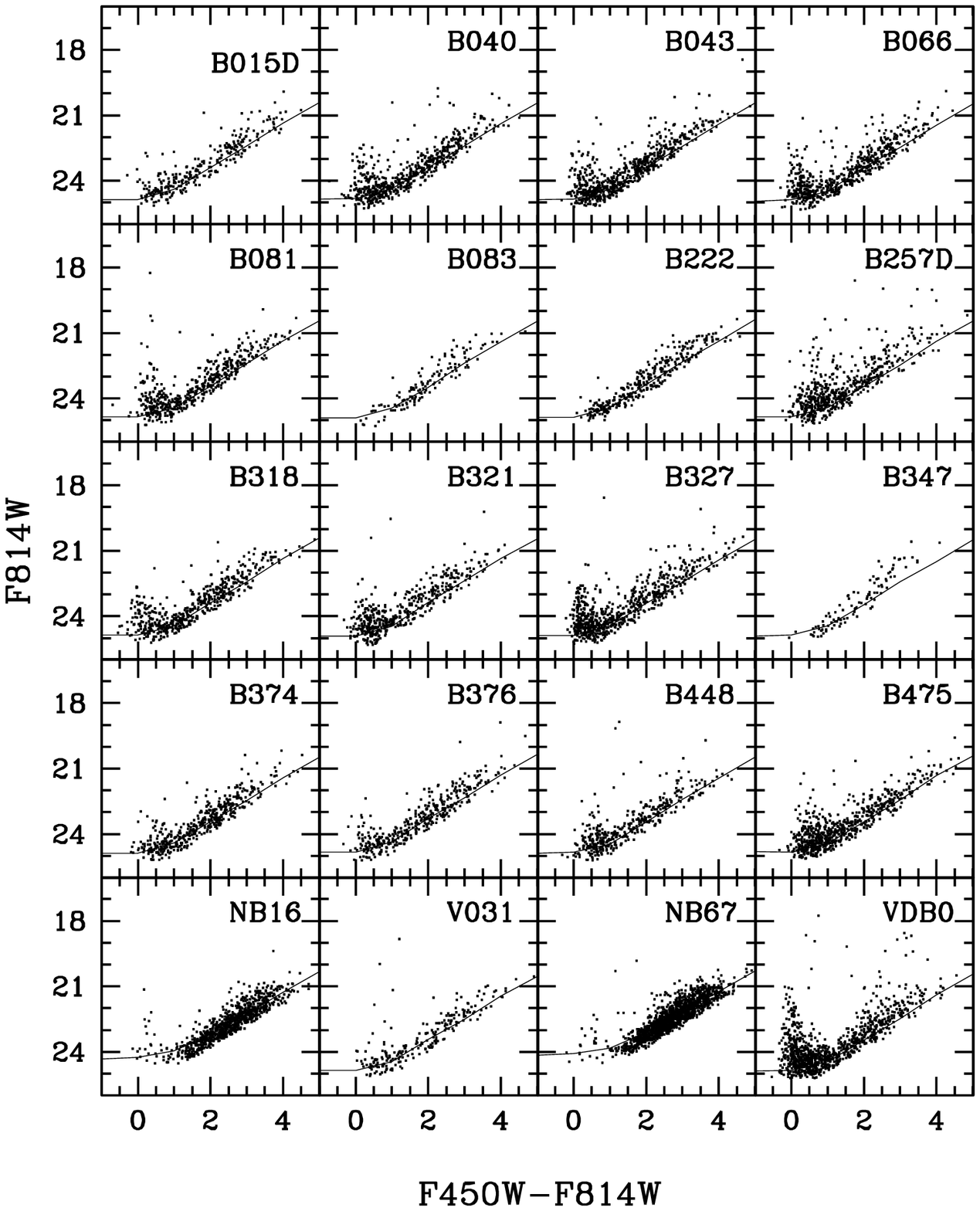}
      \caption{CMDs of the fields surrounding the target clusters.
       Only stars lying in the radial range $5\arcsec r\le16.5\arcsec$
       on the PC chips are plotted. The thin lines are the loci where the
       completeness reaches 50\%.
      }
         \label{cmds}
   \end{figure*}
%
%

In the following we will analyze only the CMDs of the radially selected samples,
as the best representation of the population of each cluster. The CMDs of the
surrounding fields are shown in Fig.~\ref{cmds}, for comparison with those of
the respective clusters that are studied in detail in Sect.~\ref{secabe}.

Fig~\ref{prof1} and \ref{prof2} deserve some further comment. First of all, it
has to be noted that all the clusters (at their centers) show an overdensity of
a factor of $\ga 10$ with respect to the surrounding field, at least in one of
the two profiles. The only exception is NB16 that is so compact that only a tiny
corona is resolved into stars, resulting in a low ($\sim 2\times$) 
overdensity of red stars (but see the light profile obtained in Pap-III). Note that in many cases, the very central region of
the cluster is not fully resolved, thus the reported central overdensities are just
lower limits to the true ones. Second, there are five clusters that show no
sign of overdensity in the Blue Box. B083, B347, and NB16 have been discussed
above;  they cannot be considered as YMC candidates anymore. B222
and B374 on the other hand have both $H_{\beta}>3.5\AA$. In four cases the
cluster show no sign of overdensity in the Red Box, in particular, B040, B043,
B066, B327. In all the other cases, the overdensity is detected in both the Blue
and Red boxes populations, even if not necessarily in similar degree. In general
the overdensity from MS stars is larger than in RGB/AGB/RSG, as expected from
evolutionary considerations (Renzini \& Fusi Pecci \cite{rfp}).


\section{Age and metallicity}
\label{secabe}

Once established that our targets are real clusters, the main purpose of our
survey is to obtain a reliable age estimate for all of them from their CMDs. 
This
will be done by comparison with theoretical isochrones from the set by Girardi
et al. (\cite{gir02}, G02 hereafter, the models are in the same photometric
system as the data; see Pap-I for a discussion about the choice of the set of theoretical models), following the approach described in detail
in Pap-I. The procedure provides a simultaneous estimate of the age, the
reddening and the metallicity of each cluster under consideration, by 
eye-aided isochrone fitting. In Pap-I we have shown that the data from our survey can be
used to reliably estimate ages in the range from $\sim 10$ Myr to $<500$ Myr
(also depending on the total mass of the considered clusters, i.e. on the number
of stars populating the MS),
from the luminosity and color of the Turn Off (TO) point. The distribution of
RSG may help to constrain the metallicity of the population, while the color of 
the blue edge of the MS is the best indicator of the degree of interstellar 
extinction (see Pap-I).

In our sample, there are eleven clusters 
that have a significant number of MS
stars brighter than $F814W=24.0$. As the completeness of the sample is $C_f\ga
80$\% above this limit, (in the color range enclosing the MS, see Fig.~\ref{compl}), reliable completeness-corrected 
LFs of the MS population can be obtained, and used to
further constrain the age of these clusters, as one in Pap-I. All of these
eleven clusters have ages lower than $\simeq 200$ Myr. They are homogeneously
analyzed in Sect.~\ref{primi}. Also VdB0 belongs to this class but it is not
considered here as it has been already treated in Pap-I.
 
Two clusters (B475 and V031) show a clear MS population only for 
$F814W>24.0$. As their observed MS lie in a range where the completeness factor
drops from $C_f\sim$ 80\% to $C_f\sim 0$ in $\sim 2$ magnitudes their LF
would be strongly affected by large completeness corrections. For these reason
we limit our analysis to isochrone fitting for these clusters 
(Sect.~\ref{secondi}).

Finally, there are five clusters that do not display any obvious MS population
in the range of magnitudes accessible with our data. For these clusters we can
provide only a strong lower limit to their age, that must be older than
300-500 Myr. These clusters are discussed in Sect.~\ref{terzi}.
The final results
of the analysis of the CMD presented below are reported in Tab.~\ref{table:2}.

\subsection{Clusters with bright MS (age$<$ 200 Myr)}
\label{primi}

   \begin{figure*}
   \centering
   \includegraphics[width=\textwidth]{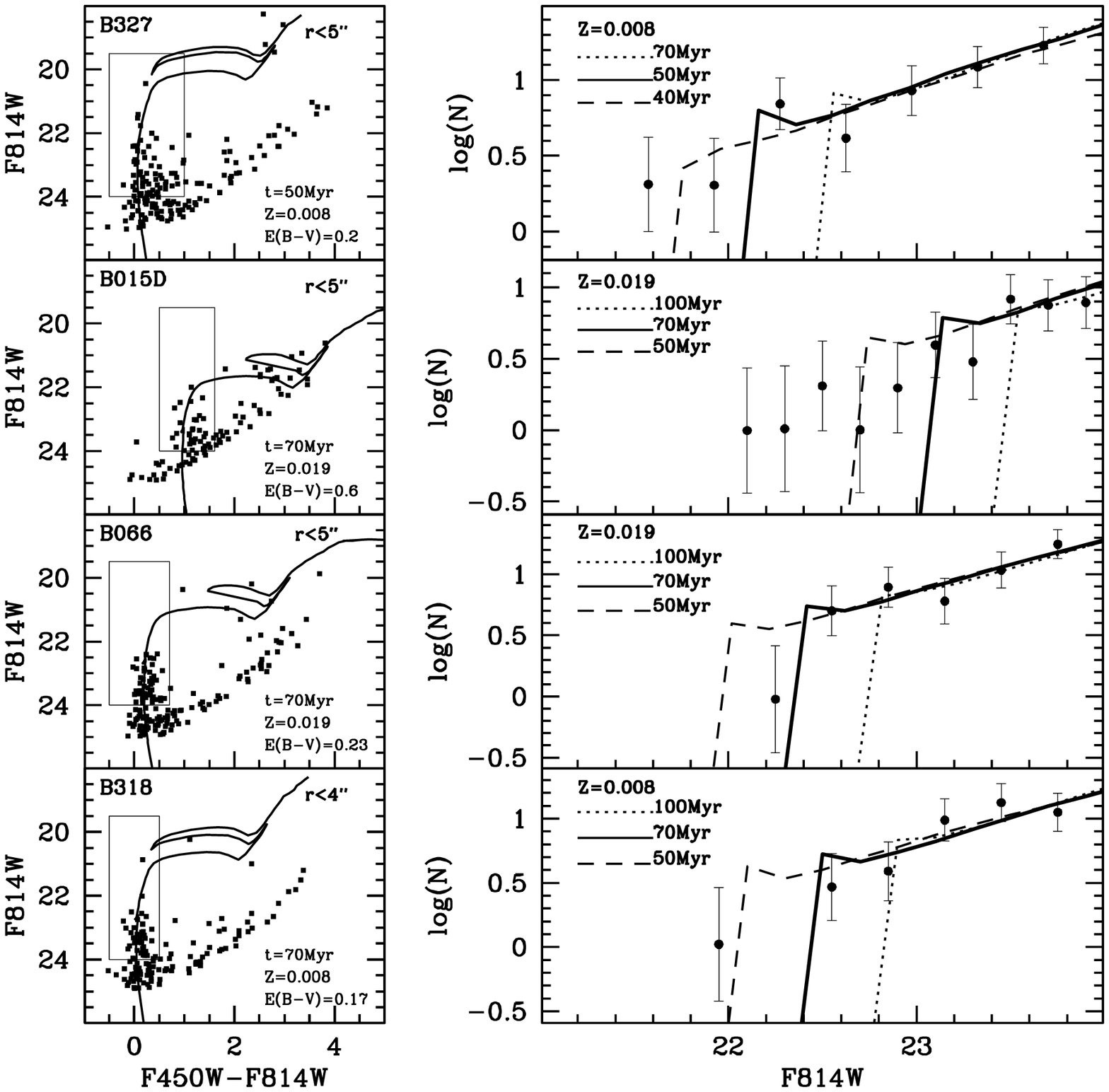}
      \caption{Left panels: CMDs of the clusters B327, 
      B015D, B066, and B318, displaying only stars within the radial selection 
      reported in the upper right corner of each panel. 
      The adopted best-fit value of the reddening and the age and metallicity of
      the best-fit isochrone (thick continuous line) are reported in the lower
      right corner of each panel. The rectangular boxes adopted
      to select the stars used to obtain the LFs shown in the
      right panels are also plotted.      
      Right panels: the observed completeness-corrected LFs of the cluster MS
      (filled circles with error bars)
      are compared with theoretical models of different ages.
      The thick continuous line corresponds to the best-fit model shown in the
      CDMs. In all cases, it provides a reasonable fit to the observed LF and, in
      particular, to the sudden drop of star counts at the upper limit of the
      MS. The dotted and dashed lines are theoretical LFs corresponding to
      strong upper and lower limits to the age, respectively, as they are the
      nearest models that can be clearly excluded by the data.
      The theoretical LFs have been arbitrarily normalized to best match the
      three faintest observed points.}
         \label{age1}
   \end{figure*}

   \begin{figure*}
   \centering
   \includegraphics[width=\textwidth]{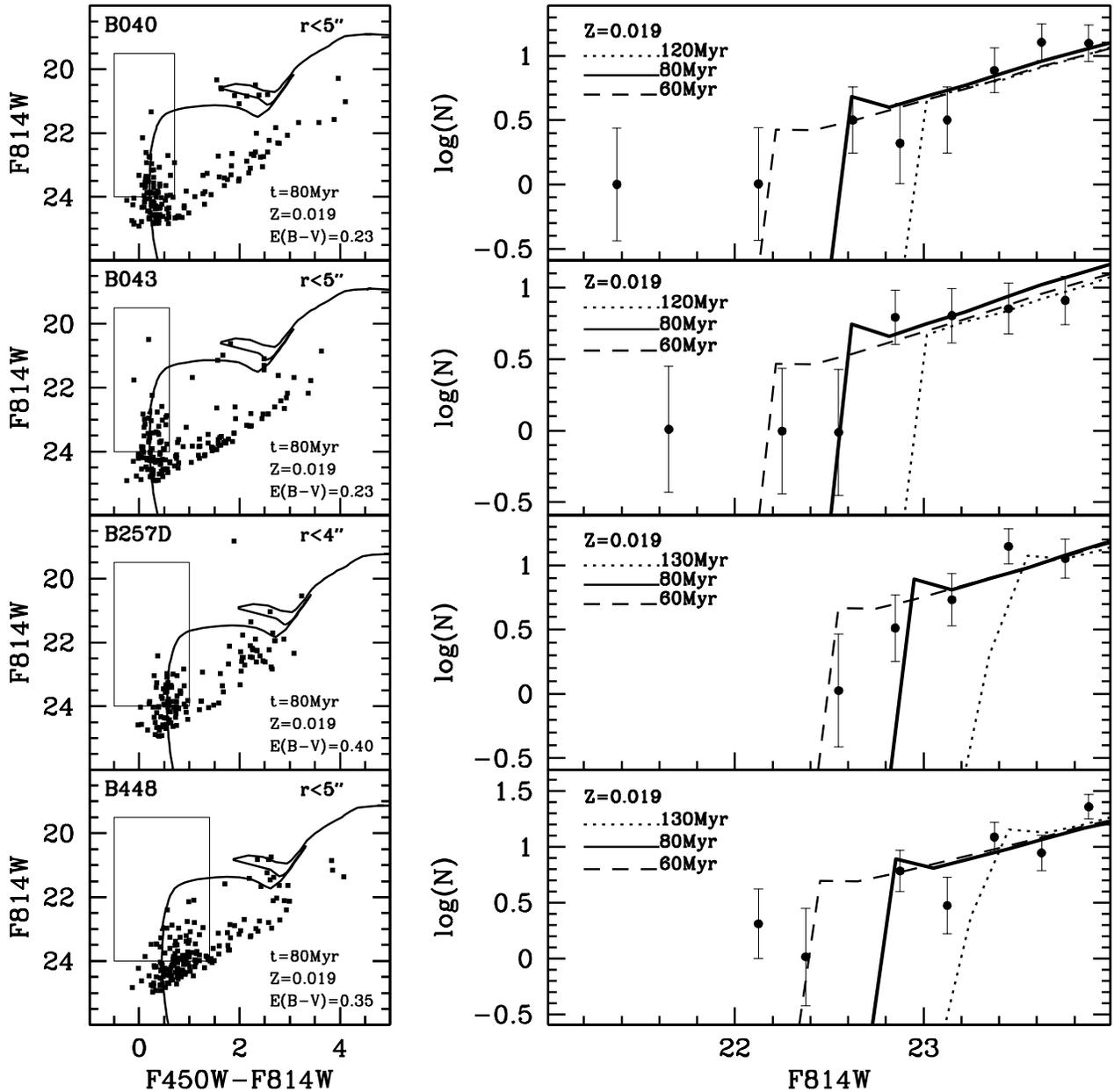}
      \caption{Same as Fig.~\ref{age1} but for the clusters B040, B043, 
      B257D, and B448.}
         \label{age2}
   \end{figure*}

   \begin{figure*}
   \centering
   \includegraphics[width=\textwidth]{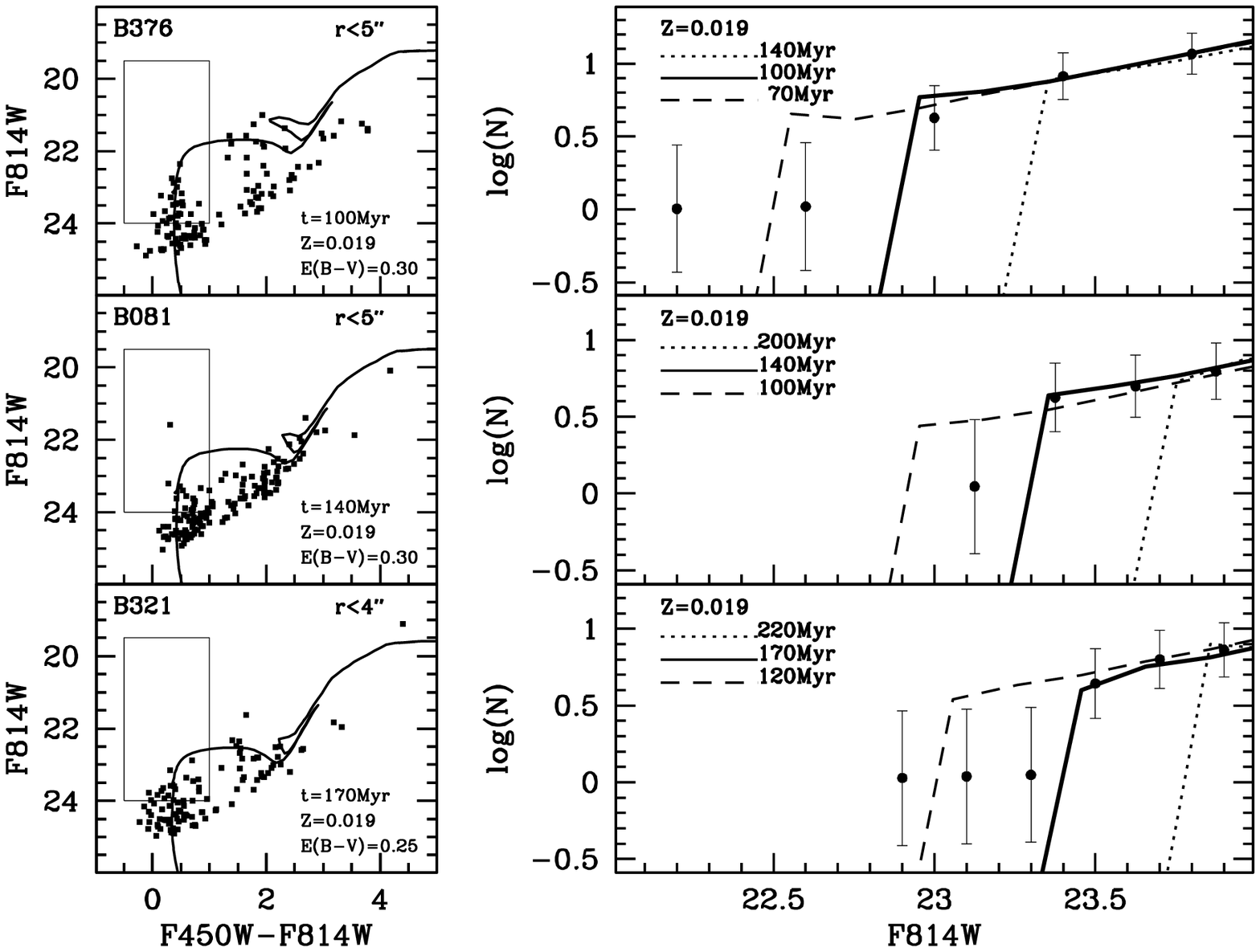}
      \caption{Same as Fig.~\ref{age1} but for the clusters B376, B081, 
      and B321.}
         \label{age3}
   \end{figure*}

Fig.~\ref{age1}, \ref{age2} and \ref{age3} show the observed CMDs and LFs of
the eleven clusters having a significant MS population brighter than 
$F814W=24.0$. The boxes overplotted on the CMDs have been used to select the
stars that were used to derive the LFs. 

For each cluster we explored the space of parameters to find the isochrone
and the reddening providing the best overall fit to the observed CMDs. As
differential reddening may move stars toward the red and the presence of 
binary
systems also has the effect of broadening the MS toward the red side, we
searched for solutions where the theoretical MS fits the blue side of the MS.
As noted above, the distribution of RSGs was used as a guide to fix the metallicity of
the best-fit model (see Pap-I). Following the approach of Pap-I, we adopt
Z=0.019 as the starting guess for the metallicity of the cluster, trying other
metallicity only if this was required to better fit some feature of the CMD.
A correct interpretation of the cluster CMD was
aided by a comparison with the CMD of the surrounding field, to establish, for
example, if a population of a few RSG can be considered as characteristic of the
cluster or compatible with belonging to the field. The typical uncertainty on
the reddening estimate is $\pm 0.04$ mag (see Pap-I).

The theoretical LF of the isochrone that best-fits the observed CMD morphology
(thick continuous line in the right panels) is compared to the observed LF
(filled dots with error bars) to check the compatibility of the solution with the
star counts (Salpeter's \cite{salp} Initial Mass Function is adopted).
In all the cases considered  the adopted theoretical LF is in good agreement 
with
the observations and, in particular, it reproduces the sudden drop in star
counts corresponding to the upper luminosity limit of the MS, a feature that is
mainly sensitive to age (see Pap-I and references therein). Two theoretical LFs
of the same metallicity as the main solution but different ages are used to
show the maximum and minimum age that are not compatible with the observed LF.
The difference between these values and the age of the best-fit solution 
are taken as the uncertainty associated with our age estimate. 
Nine of the eleven clusters considered in this section have ages 
between 50 Myr and 100 Myr. All of them show a recognizable (and in same case
sizable, see B040, for example) population of RSG stars, in addition to an obvious MS.
The other two clusters, B081 and B321 have ages of 140 and 170 Myr,
respectively.

\subsection{Clusters with faint MS (200 Myr$\le$age$\le 500$ Myr)}
\label{secondi}

\begin{figure*}
   \centering
   \includegraphics[width=15cm]{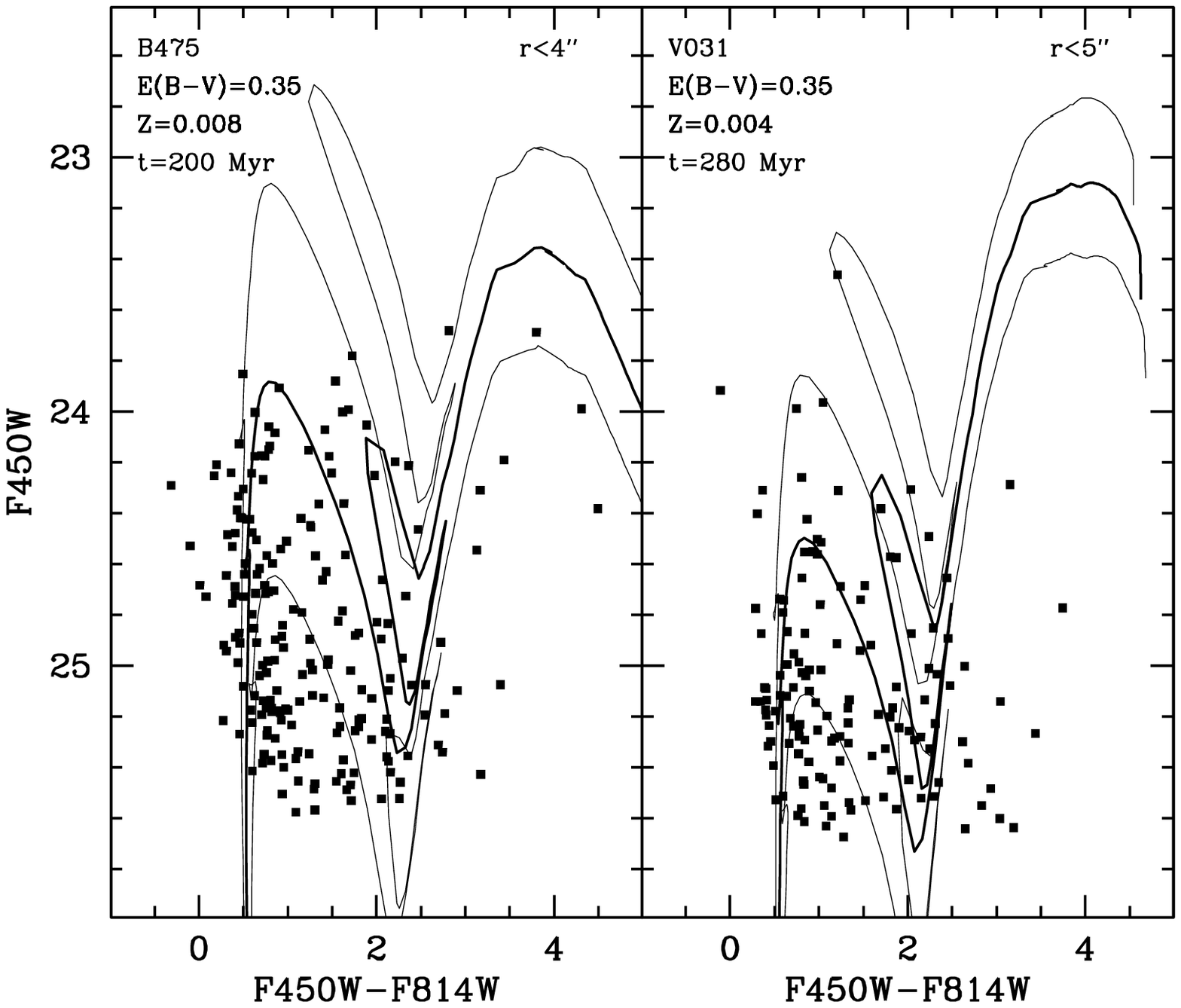}
      \caption{Observed CMDs of the clusters B475 (left panel) and V031 (right
      panel) in the plane F450W vs. F450W-F814W where the MS population of 
      these older clusters is more clearly visible. Only stars with the radial 
      selection reported in each panel are plotted. The best-fit isochrone is 
      plotted as  thick line (age, metallicity and reddening values are 
      reported in each panel). The thin isochrones bracket the upper and 
      lower limits on the age, and correspond to age $\simeq$ 125~Myr and 
      315~Myr for
      B475, and age 200 Myr and 400 Myr for V031.}
         \label{ageB}
   \end{figure*}

Fig.~\ref{ageB} shows the CMDs of the two clusters whose MS is fainter than 
$F814W=24.0$. The F450W magnitude is plotted here instead of F814W (adopted in
Fig.~\ref{age1}, \ref{age2} and \ref{age3}) as this makes the faint MS of these clusters more
clearly visible. The best fit isochrones are plotted as thick lines. The thin
lines are isochrones having ages that bracket the age solutions that can be
considered still compatible with the data. The difference in age between these
solutions and the assumed best-fit are adopted as the
uncertainty associated with our age estimates for this cases (see Pap-I). 
The two clusters have ages of $\simeq$200 Myr (B475) and $\simeq$280 Myr 
(V031).

%
%

%
   \begin{figure*}
   \centering
   \includegraphics[width=15cm]{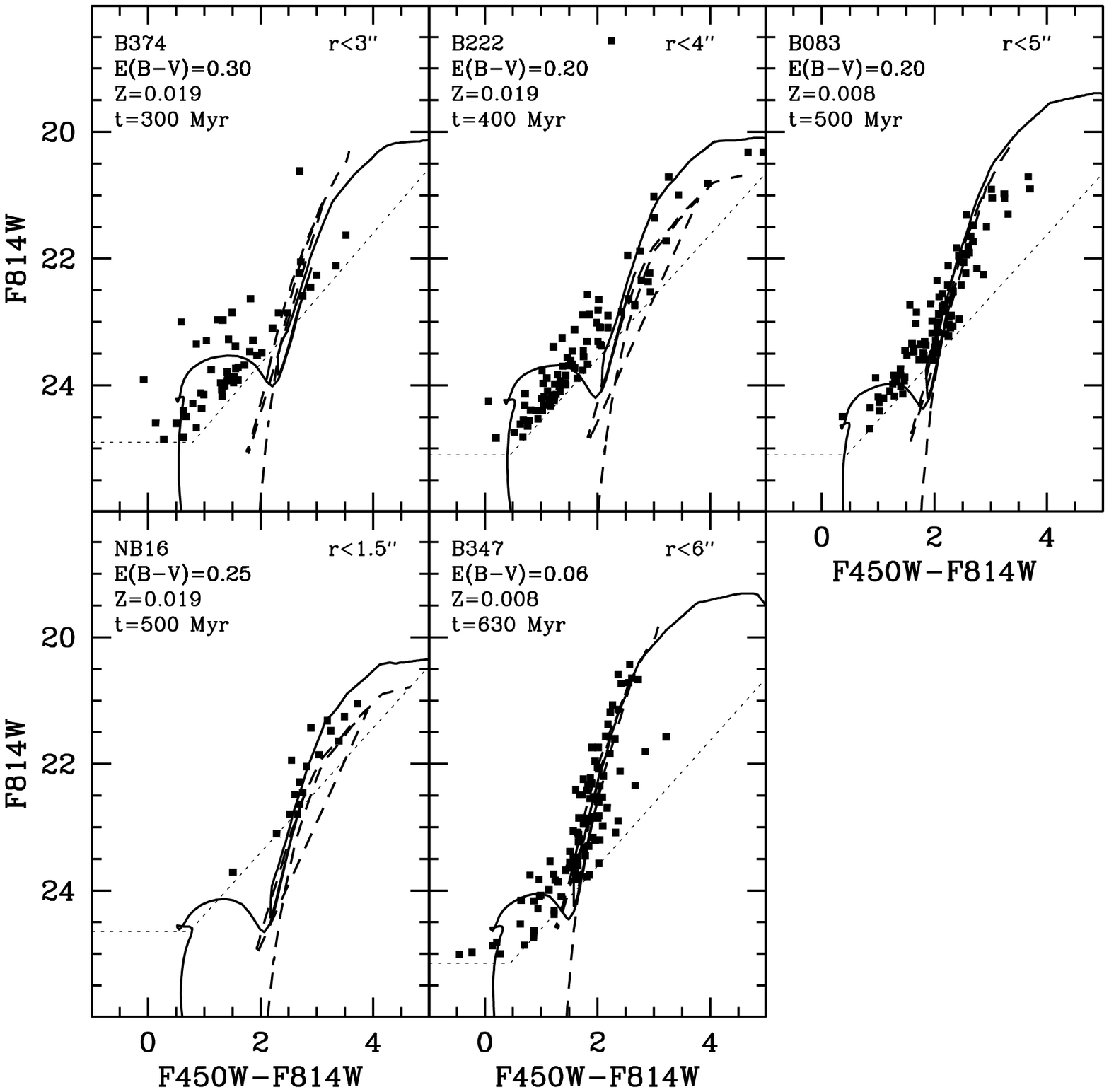}
      \caption{CMDs of the clusters B374, B222, B083, NB16, and B347. 
      Only stars within the radial selection reported in each panel are plotted.
      The thin dashed lines marks the locus where the completeness of the sample
      reaches $\simeq$ 0\% (see Pap 1), to illustrate the selection effects on
      the CMD morphology imposed by the run of limiting magnitude as a function
      of color.  In each panel, the continuous line is the youngest age
      isochrone that is compatible with the observed CMD, providing a
      strong lower limit to the age of each cluster. The adopted age, metallicity and
      reddening values are reported in the upper left corner. 
      The dashed line is a 12 Gyr old isochrone matching the color of the
      observed RGB. The metallicity of these old-age 
      isochrones is $Z=$ 0.001, 0.004, 0.001, 0.004, and 0.001  for
      B374, B222, B083, NB16, and B347, respectively.}
         \label{ageI}
   \end{figure*}
%

\subsection{Clusters whose MS is not detected (age$> 500$ Myr)}
\label{terzi}

Fig.~\ref{ageI} shows the CMDs of the clusters that do not display a clear MS in
the considered range of magnitudes. In each panel we plot (a) the ``youngest'' 
isochrone that is compatible with the observed CMD morphology, to provide a firm
lower limit to the age of these clusters (thick continuous line), and, (b) a 12
Gyr old isochrone (thick dashed line), showing that the observed CMD is also
compatible with very old ages. In all the cases we adopt the metallicity value
that provided a satisfactory match of the color of the (putative) RGB. 

Three of the five clusters considered here (B083, NB16 and B347) have
integrated properties that are compatible with old ages 
(see Sect.~\ref{sample}). B083 and B347 display a steep and well populated red
sequence, much bluer than the limits imposed by the run of the completeness as a
function of color (thin dotted lines), typical of the RGB of classical old (and metal deficient) GCs.
The handful of stars resolved in NB16 are also compatible with being near the tip of an old RGB, but their scarcity poses strong caveats on any interpretation.

B347 and B222 are more interesting cases: both have two independent 
concordant estimates of
$H_{\beta}$  indicating $H_{\beta}> 4.0 \AA$, and both have some stars
just above the detection limits in the blue, that may be compatible with the bright end of a fainter MS. The observational scenario is fully consistent with
the hypothesis that these two clusters might be intermediate-age (age$\sim 0.5 - 2$
Gyr). A deeper photometry follow-up is clearly required to settle the 
issue of the age of these clusters. It is worth noting that a convincing 
case for an M31 cluster in the age range 1-8 Gyr with age estimated from
a CMD has never been provided.

\section{Masses from ages and J,H,K integrated photometry}
\label{massageir}

In Table~\ref{table:2} we report the age, metallicity and reddening estimates
obtained from the analysis of the CMDs presented above. To increase the sample of
YMC to be considered in the following we added a total of 10 further clusters
whose ages have been derived from CMDs obtained from HST data in a way
fully homogeneous with that adopted here. In particular we add six clusters from
Perina et al. (2009b, P09b hereafter) and four clusters from 
Williams \& Hodge (2001, WH01 hereafter; see Pap-I). All of them lie in the range of
V luminosities typical of YMC ($M_V\la -6.5$, according to F05), with the only
(possible) exceptions of M050 and M039 that appear somewhat fainter than this, and of B521 that lacks an estimate of its V magnitude (but it is found to have a mass similar to other YMC, based on its Near Infrared Magnitudes, see below). 
We decided to keep these clusters within our sample, being well 
aware that the threshold between the brightest of the clusters studied in 
Pap-II and Krienke \& Hodge (~\cite{krie1,krie2}) and the faintest 
clusters considered here is 
somewhat blurred, both by lack of a clear-cut definition and by 
observational uncertainties. In particular, Fig.~\ref{agemass}, will 
show that some of the clusters studied in Pap-II appear to have 
masses typical of YMC. Still we preferred not to include these massive 
Pap-II clusters as main objects of the present analysis as most 
of them have their ages estimated from integrated colors, i.e. with significantly greater uncertainties than those obtained here from CMDs (see, e.g., Fig.~8 of Pap-II)\footnote{There are only  two clusters from Pap-II having $M_V\la -6.5$ and ages estimated from their CMD, but also in these cases the associated age 
uncertainties are relatively large, i.e. 0.5-0.6 dex in log(Age) vs. a typical uncertainty of 0.2 dex for our main sample, see Tab.~\ref{table:2}.}. 

Five of the newly included clusters are projected onto the {10 kpc ring}, as most of our original targets, four lie slightly nearer to the center of the galaxy, and one is in the outskirts of the visible disk (see Fig.~\ref{cpos}). 
B049, B367, B458, B315 and B317 have two independent estimates of $H_{\beta}$, all 
of them higher than $4.5\AA$ (F05, G09). B342 has just one estimate
($H_{\beta}=7.06\AA$, FP05), while the other four clusters lack any measure of
this index. B368 lacks $H_{\beta}$ but has $(B-V)_0=0.06$. For M039, M050 and
B521  there is no $(B-V)_0$ estimate available. In any case all the six clusters
from P09b and the four from WH01 have age $<1$ Gyr, as derived from their CMD.

\begin{table*} 
\centering
\caption{Newly derived ages, metallicity and reddening for the target clusters 
 and other clusters included in the analysis$^a$.}
\label{table:2}
  \begin{tabular}  {@{}lcccccc@{}}
  \hline \hline
Name  &    log(t) & $\Delta$log(t)  & Z  & E(B-V) & M$_v$$^b$ \\ 
 \hline
 This survey\\ 
 \hline 
 \\   
B015D-D041   &  7.85   & $\pm0.15$          & 0.019 & 0.60 & -8.53 \\ [1ex] 
B040-G102    &  7.90   & $_{-0.15}^{+0.20}$ & 0.019 & 0.23 & -7.80 \\ [1ex] 
B043-G106    &  7.90   & $_{-0.15}^{+0.20}$ & 0.019 & 0.23 & -8.22 \\ [1ex] 
B066-G128    &  7.85   & $\pm0.15$          & 0.019 & 0.23 & -7.76 \\ [1ex] 
B081-G142    &  8.15   & $\pm0.15$          & 0.019 & 0.30 & -8.60 \\ [1ex] 
B257D-D073   &  7.90   & $_{-0.15}^{+0.20}$ & 0.019 & 0.40 & -8.31 \\ [1ex] 
B318-G042    &  7.85   & $\pm0.15$          & 0.008 & 0.17 & -7.98 \\ [1ex] 
B321-G046    &  8.23   & $_{-0.15}^{+0.10}$ & 0.019 & 0.25 & -7.57 \\ [1ex] 
B327-G053    &  7.70   & $_{-0.10}^{+0.15}$ & 0.008 & 0.20 & -8.51 \\ [1ex] 
B376-G309    &  8.00   & $\pm0.15$          & 0.019 & 0.30 & -7.34 \\ [1ex] 
B448-D035    &  7.90   & $_{-0.15}^{+0.20}$ & 0.019 & 0.35 & -8.07 \\ [1ex] 
B475-V128    &  8.30   & $\pm0.20$          & 0.008 & 0.35 & -8.00 \\ [1ex] 
V031         &  8.45   & $\pm0.15$          & 0.004 & 0.35 & -8.12 \\ [1ex] 
VDB0         &  7.40   & $\pm0.30$          & 0.019 & 0.20 & -10.03\\ [1ex] 
B083-G146    & $>$8.70 & $\dots$ 	     & 0.008 & 0.20 & -8.00 \\ [1ex] 
B222-G277    & $>$8.60 & $\dots$ 	     & 0.019 & 0.20 & -7.66 \\ [1ex] 
B347-G154    & $>$8.80 & $\dots$ 	     & 0.008 & 0.06 & -8.16 \\ [1ex] 
B374-G306    & $>$8.50 & $\dots$             & 0.019 & 0.30 & -7.09\\ [1ex] 
NB16         & $>$8.70 & $\dots$ 	     & 0.019 & 0.25 & -7.69 \\ [1ex] 
 \hline
P09b\\
 \hline
\\
B049-G112    &  8.45   & $\pm0.20$ 	    & 0.019 & 0.30 & -7.84  \\ [1ex] 
B367-G292    &  8.30   & $\pm0.20$ 	    & 0.019 & 0.25 & -6.79  \\ [1ex] 
B458-D049    &  8.50   & $\pm0.20$ 	    & 0.019 & 0.25 & -7.40  \\ [1ex] 
B521         &  8.60   & $\pm0.30$ 	    & 0.019 & 0.55 & $\dots$ \\ [1ex] 
M039         &  8.50   & $\pm0.20$ 	    & 0.019 & 0.10 & -5.84  \\ [1ex] 
M050         &  8.75   & $\pm0.30$          & 0.019 & 0.15 & -6.22   \\ [1ex] 
\hline
WH01\\
\hline                                                                      
\\
B315-G038     &  8.00   & $_{-0.20}^{+0.15}$ & 0.008 & 0.31 & -8.96 \\ [1ex] 
B319-G044     &  8.00   & $_{-0.20}^{+0.15}$ & 0.008 & 0.23 & -7.57  \\ [1ex] 
B342-G094     &  8.20   & $_{-0.20}^{+0.15}$ & 0.008 & 0.20 & -7.36  \\ [1ex] 
B368-G293     &  7.80   & $\pm0.10$          & 0.019 & 0.20 & -7.17  \\ [1ex] 
\hline                                                        
\end{tabular} 
                                                       
\begin{list}{}{}
\item For five surveyed clusters only a lower limit to the age can be obtained 
from our CMDs.  
\item[$^{\mathrm{a}}$] The additional clusters are six clusters studied in 
Perina et al (2009a), from HST archive data, and the four clusters studied by 
Williams \& Hodge (2001).
\item[$^{\mathrm{b}}$] Integrated V magnitudes from the RBC.  
\end{list}

\end{table*} 

To derive the most reliable estimate of the total stellar mass of the clusters
in our sample we couple our age estimates with integrated Near Infra Red (NIR)
photometry, as stellar mass-to-light ratios in NIR bands have a much shallower dependence on age than their optical counterparts (see Pap-I for discussion). As the best estimate of the integrated
J,H,K magnitudes we took the values of the $r=10\arcsec$ aperture magnitudes 
from the 2MASS-6X-PSC catalog (see Nantais et al.~\cite{nantais}), that is
obtained from deeper observations (with respect to the normal 2MASS data, Skrutskie et al.~\cite{skrut}) over a limited region of the sky that,
luckily, includes M31. The adopted NIR photometry as well as the accurate
positions reported in 2MASS-6X-PSC are listed in Table~\ref{table:3}. Only two
clusters have no valid measures in 2MASS-6X-PSC, i.e. B367 and M039. To preserve the homogeneity of the analysis we do not include these clusters in any of the following analyses that make use of mass estimates, however, for completeness, in Tab.~\ref{table:3} we provide a tentative mass estimate derived from the log(age) $vs.$ $M_V$ diagram presented in Fig.~\ref{Mvage}.
The
apparent magnitudes are transformed into absolute ones adopting the reddening
estimates derived here (Tab.~\ref{table:2}), the distance modulus
(from McConnachie et al.~\cite{dista}) and the reddening laws 
(from Rieke \& Lebofsky~\cite{rieke}) adopted in Pap-I.

   \begin{figure*}
   \centering
   \includegraphics[width=16cm]{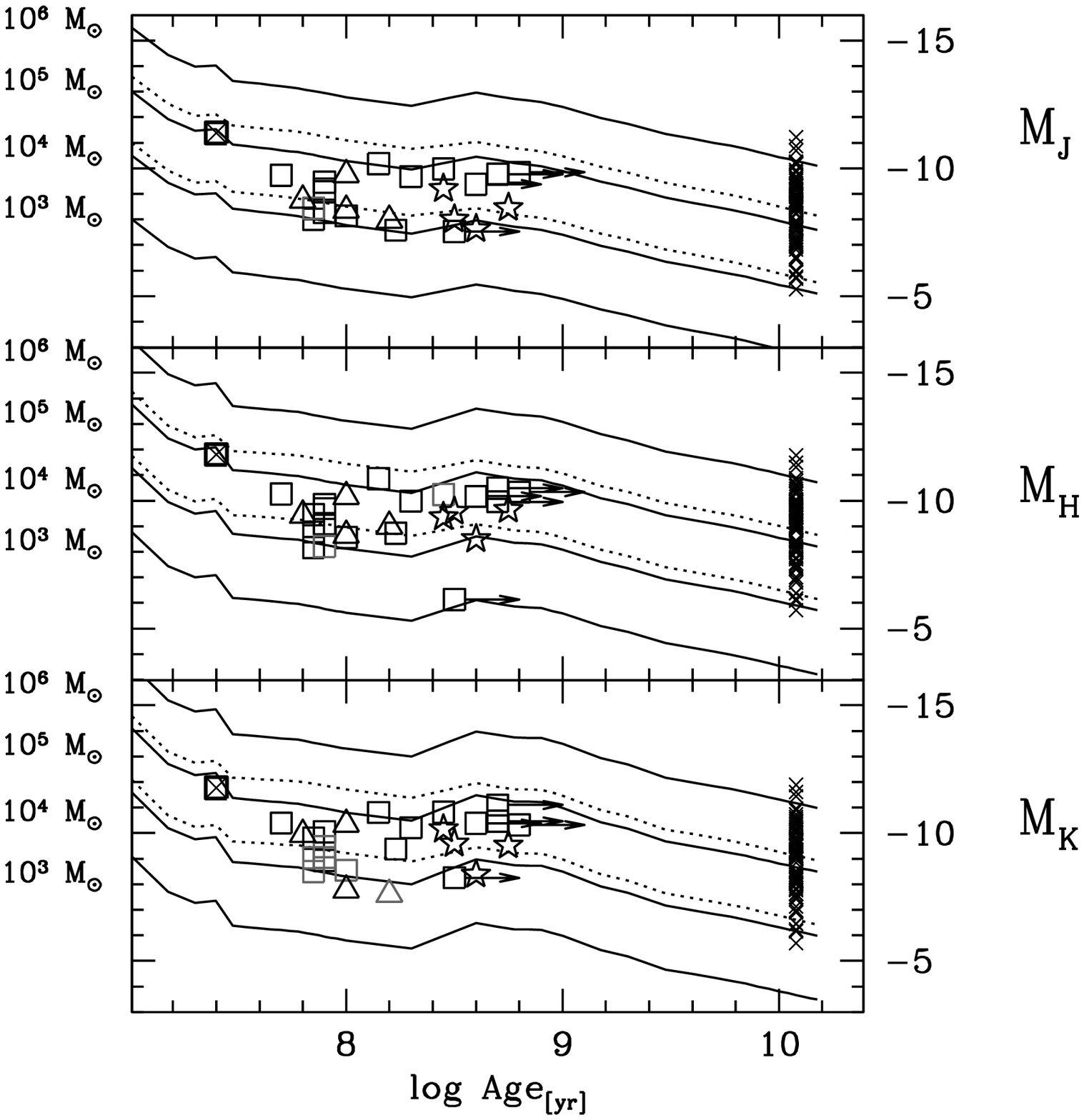}
      \caption{Log(age) vs. integrated magnitude plane for 
      near infrared colors. The target clusters are represented as open squares (VDB0 as a crossed
      square), the clusters from P09b as open stars, and the clusters from WH01 
      clusters as open triangles,
      IR magnitudes are taken from Tab.~\ref{table:3}.     
      Note that B367 and M039 are not plotted because they lack
      NIR photometry in the 2MASS-6X-PSC catalog.
      The gray symbols show the clusters that have "null" error on IR magnitudes
      in the 2MASS-6X-PSC catalog.
      Integrated magnitudes of Galactic GCs ($\times$ symbols) 
      are taken from Cohen et al. (\cite{cohir}).
      The continuous lines are fixed-stellar-mass models from the set
      by Maraston (1998, 2005) for SSPs of solar metallicity, with a
      Salpeter's Initial Mass Function (IMF) and intermediate Horizontal Branch
      morphology. Note that in this plane, the dependence of the models from 
      the assumed IMF, metallicity and HB morphology is quite small 
      (see B08).       
      The dotted lines are $M=10^4 M_{\sun}$ and $M=10^5 M_{\sun}$
      iso-mass models assuming a Kroupa \cite{kroupa} IMF instead of a 
      Salpeter (\cite{salp}) IMF, plotted here to illustrate the weak effect 
      of assumptions on IMFs.}
         \label{Mir}
   \end{figure*}
%

In Fig.~\ref{Mir} we compare the position of our clusters in the integrated
(J,H,K) magnitude vs. log(age) plane with a grid of models of Simple Stellar
Population (SSP) of solar metallicity and various total mass, from the set by 
Maraston (\cite{mara1}, \cite{mara2}, see Pap-I). In B08 and in
Pap-I we have shown that the mass that can be deduced from these plots depends only 
weakly on the assumed metallicity and IMF. Here we get an independent estimate
of the mass from each (J,H,K) plot and we take the weighted average of the three values as our final estimate. The uncertainties were obtained on each individual estimate from J, H, K by
finding the maximum interval in mass that was compatible with the errors in age
and in integrated magnitudes. Then the three values (per cluster) were combined into the final {\em weighted} error that is reported in Table~\ref{table:3} together with the final mass estimates. 

It is very reassuring to note that the three plots provide very similar age
estimates: all the clusters  considered appear to have masses between $\sim 10^4
M_{\sun}$ and $\sim 10^5 M_{\sun}$. The estimates from the three different 
NIR magnitudes typically agree within a factor of 2. The adoption of a 
Kroupa (\cite{kroupa}) IMF
instead of that of  Salpeter would change the mass estimates by less than a factor of
2 (Pap-I). The adoption of different sets of models would lead to a maximum difference of the same amount in the final mass estimates (we have compared the $M/L$ predictions adopted here with those from the sets by Pietrinferni et al.~\cite{basti} and Bruzual \& Charlot~\cite{bruz}, in the age range that is relevant for our clusters). Finally, if models with age-dependent $M/L$ are adopted (i.e. including the effects of differential mass loss, Kruijissen\& Lamers~\cite{kruij}), the mass estimates for our clusters change by a mere $\la 20\%$
(see also Pap-III).
Taking all of these factors into account it turns out that our mass estimates should be accurate within a factor of $\la 3$, as confirmed also by the comparison with the independent estimates from Pap-III and C09.

There is only one case of significant disagreement in the position of a cluster 
in the different NIR passbands, i.e. B347 whose reported
H magnitude implies a (lower limit) mass estimate nearly one order of magnitude
lower than J and K. We attribute this occurrence to an error of the integrated H magnitude reported in 2MASS-6X as this value is at 
odds with that of all the other clusters while B347 is normal 
in all other respects. For instance it has a J-K color well within the range of the other clusters of the sample while its H-K color is more than one magnitude redder than any other.
Finally we note that the independent lower limit mass  obtained from the 
log(age) vs. $M_V$ diagram (see Fig.~\ref{Mvage}), 
are in good agreement with that estimated from J and K magnitude for B347. 
Finally, as we have obtained just a lower limit to the age of B347 we do not provide 
an age estimate for this cluster. 
B347 as well as all the other clusters for which we can provide only a lower limit to the age are not included in the analysis of Sect.~\ref{disc} that is limited 
to the young clusters that constitute the main subject of our study.


   \begin{figure}
   \centering
   \includegraphics[width=9cm]{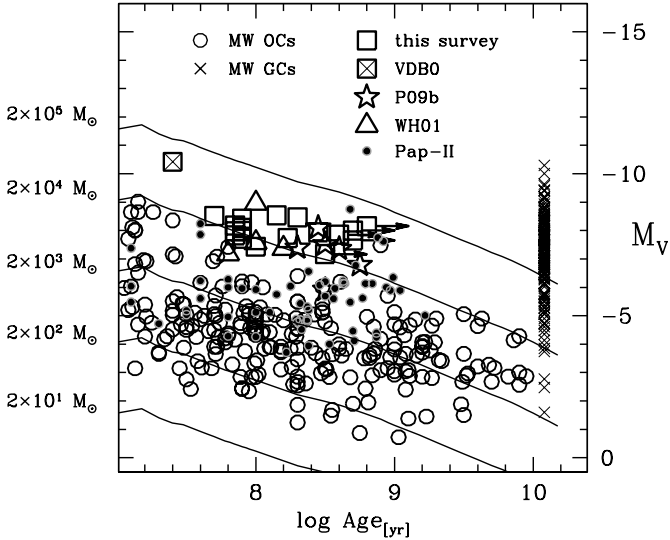}
      \caption{Integrated V mag and total mass as a function of age for 
      various samples of clusters. Galactic open clusters (OC, from the 
      WEBDA database) are plotted as filled circles, 
      Galactic globular clusters (GC, $M_V$ from the most recent version of the 
      Harris (1996) catalog, i.e. that of February 2003, the ages have been
      arbitrarily assumed to be 12.0~Gyr for all the clusters) are plotted as
      $\times$ symbols.      
      The target clusters are represented as open squares (VDB0 as a crossed
      square), the clusters from P09b as open stars, and the clusters from WH01 
      clusters as open triangles. 
      $M_V$ magnitudes of the target clusters and of the P09b clusters are from the 
      new aperture photometry performed on the CCD images by Massey et al. (2006),
      except for B083 and B347
      whose magnitudes are from RBC (see Tab.~\ref{table:1}. 
      $M_V$ magnitudes of the WH01's clusters are from RBC. Log Age is from 
      Tab.~\ref{table:2}. Points with arrows have only lower limits to the age. 
      Filled circles are M31 OCs from Pap-II.
      The continuous lines are fixed-stellar-mass models from the set
      by Maraston (1998, 2005) for SSPs of solar metallicity, with a
      Salpeter's Initial Mass Function (IMF) and intermediate Horizontal Branch
      morphology. Note that in this plane, the dependence of the models from 
      the assumed IMF, metallicity and HB morphology is quite small 
      (see B08). 
      The outlier OC at log Age$\simeq 9.0$ is Tombaugh~1.}
         \label{Mvage}
   \end{figure}
%

   \begin{figure}
   \centering
   \includegraphics[width=9cm]{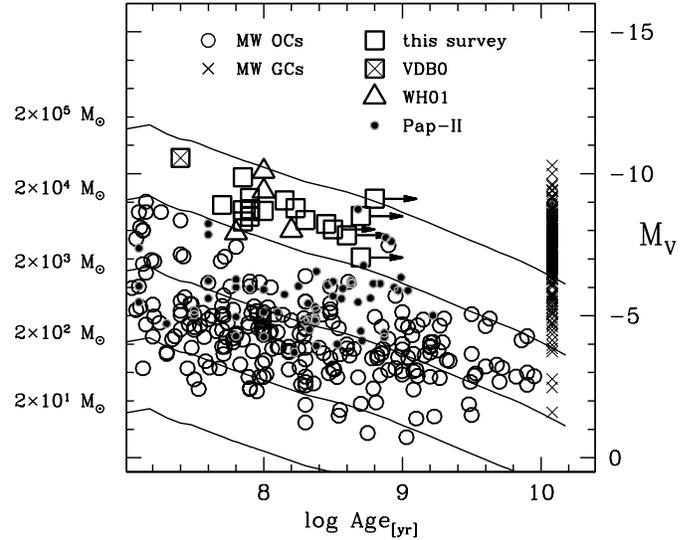}
      \caption{Same as Fig.~\ref{Mvage} but with $M_V$ magnitudes of the target 
      clusters and of the WH01's clusters obtained from fitting
      King (1966) models to our HST data, from Pap-III.
      The clusters from P09b are not included in the plot as they have not been
      considered in Pap-III.
}
         \label{MvageB}
   \end{figure}
%

\subsection{Comparison with Galactic open clusters}

In Fig.~\ref{Mvage} we show the log(age) vs. absolute magnitude plot analogous
to Fig.~\ref{Mir} but using $M_V$ instead of $M_J$, $M_H$, $M_K$. While NIR
magnitudes are preferred to get reliable estimates of the stellar mass of our
clusters (see Sect.~\ref{massageir} and Pap-I), the use of $M_V$ allows us a direct comparison with different kinds of
clusters for which integrated magnitudes in NIR passbands are lacking, 
Galactic OCs in particular (B08, Pap-I).

Inspection of Fig.~\ref{Mvage} confirms the tentative conclusions of Pap-I
(and F05). The distribution of our target clusters marginally overlaps with the
high-mass tail of the Galactic OC distributions, but {\em the bulk of the sample
of candidate YMC considered here is significantly more massive than Galactic OCs
in the same age range}. In this sense, the brightest, most massive and youngest
cluster of our sample, VdB0 having age=25 Myr and $M\simeq 6\times
10^4~M_{\sun}$, may appear similar to the handful of massive 
young clusters recently identified in the Milky Way (see Figer \cite{figer} and 
Messineo et al. \cite{maria}, hereafter M09, for recent reviews), 
that have masses between $0.7\times 10^4~M_{\sun}$ and 
$4.0\times 10^4~M_{\sun}$ and ages between 0.3 Myr and 18 Myr, according to
M09. The other clusters of our sample have similar (or slightly greater) masses
than the Galactic YMC but they are all significantly older (by a factor of
$>2\times$, see Sect.~\ref{disc} for further discussion). It is worth to note that  the masses estimated from Fig.~\ref{Mvage} are in agreement with those from Fig.~\ref{Mir}, typically, within a factor of 2.

In Pap-I we showed that in the case of VdB0, an exceptionally extended
cluster, the integrated magnitudes reported in the RBC were significantly 
underestimated. 
However our shallow HST
exposures were not ideal to perform integrated photometry on such large areas
(VdB0 cover the whole extent of the PC field).
For these reasons we recurred to the new homogeneous CCD survey by Massey et al.~(\cite{massey}; see Pap-I for discussion) to obtain a reliable
estimate of the total luminosity of that cluster; as said, the integrated B,V magnitudes for the clusters considered here have been obtained from the same source and with the same method (Tab~\ref{table:1}). These cases 
are less problematic, as the clusters are more compact than VdB0. 
However, it seems wise to check how the comparisons shown in Fig.~\ref{Mvage} may
depend on the actual way in which $M_V$ is estimated. To do that we present in
Fig.~\ref{MvageB}, a new version of Fig.~\ref{Mvage} in which the $M_V$ values derived from Tab.~\ref{table:1} are replaced with $M_V$ estimates obtained in Pap-III from profile fitting (with King 1966 models) performed on our HST images 
(with the same assumptions on distance and reddening adopted here). Again, it is very reassuring to note that the conclusions drawn above from Fig.~\ref{Mvage} are fully confirmed also by the new set of $M_V$ from Pap-III. In fact, the differences between the YMC of our sample and Galactic OCs are even more pronounced in the new plot, as the total V luminosities estimated in Pap-III are larger than the values adopted here by a factor of $\simeq 1.6$, in average.
For the reasons discussed in Pap-I and for homogeneity with that analysis we retain our ground-based $M_V$ estimates as our reference. 

It is interesting to note that the clusters identified by 
Krienke \& Hodge (\cite{krie1}, \cite{krie2}), and, by analogy, 
those found in Pap-II\footnote{It should be recalled that clusters 
listed in the RBC were excluded from the analysis performed in Pap-II.}, 
have an observed LF peaking around $M_V=-3$ and virtually
 dropping to zero at $M_V\ga -6$, very similar to Galactic 
OCs (see Fig.~\ref{histo}), hence they appear as the natural counterpart of the OCs observed in the Milky Way. 

In Pap-III the problem of the survival of our target clusters was discussed 
in some detail and dissolution times including the effects of internal and 
external evolution (Lamers \& Gieles \cite{lamers}), were computed. These values are reported 
also here, in Tab.~\ref{table:3}, for convenience of the reader. The 
dissolution times of young clusters are all shorter than a Hubble time, 
hence it is likely that none of them will survive long enough to become old (age$\ga$ 10 Gyr), and some of them are probably in the latest phase of their dissolution (B321, B342; Pap-III).
However, a few clusters have dissolution times longer than 1 Gyr, and it is 
not inconceivable that some of them may reach an age of several Gyr before dissolving into the M31 disk (see Pap-III).


\begin{table*} 
\centering
\caption{Newly derived masses and dissolution times for the studied clusters.}
\label{table:3}
  \begin{tabular}  {@{}l@{}cccccccc@{}}
  \hline \hline
Name  &  $\alpha_{J2000}$ &  $\delta_{J2000}$ & J &  H  & K &  log Mass & $\varepsilon$log Mass &  $t_{diss}^{Pap-III}$\\    
& & & & & & ($M_{\odot}$)& ($M_{\odot}$) & (Myr)  \\   
 \hline 
 \\   
B015D-D041  & ~$00^h$ $41^m$ $02.74^s$ & $+41\degr$ $06\arcmin$ $36.63\arcsec$ & 17.03 $\pm$ 0.42  & 15.37 $\pm$ 0.27  & 14.89 $\pm$ 0.25  & 4.2    & 0.09  & 112      \\ [1ex]    
B040-G102   & ~$00^h$ $41^m$ $38.90^s$ & $+40\degr$ $40\arcmin$ $54.15\arcsec$ & 15.48 $\pm$ 0.08  & 14.90 $\pm$ 0.19  & 14.50 $\pm$ 0.15  & 4.6    & 0.07  & 631      \\ [1ex]    
B043-G106   & ~$00^h$ $41^m$ $42.31^s$ & $+40\degr$ $42\arcmin$ $39.86\arcsec$ & 15.58 $\pm$ 0.07  & 15.50 $\pm$ 0.31  & 15.08 $\pm$ 1.00  & 4.4    & 0.10  & 3467     \\ [1ex]    
B066-G128   & ~$00^h$ $42^m$ $03.14^s$ & $+40\degr$ $44\arcmin$ $48.55\arcsec$ & 16.25 $\pm$ 0.19  & 15.81 $\pm$ 0.47  & 16.06 $\pm$ 1.00  & 4.2    & 0.08  & 891      \\ [1ex]    
B081-G142   & ~$00^h$ $42^m$ $13.59^s$ & $+40\degr$ $48\arcmin$ $38.96\arcsec$ & 14.55 $\pm$ 0.05  & 13.77 $\pm$ 0.07  & 13.76 $\pm$ 0.06  & 5.1    & 0.04  & 955      \\ [1ex]    
B257D-D073  & ~$00^h$ $44^m$ $59.35^s$ & $+41\degr$ $54\arcmin$ $47.47\arcsec$ & 15.28 $\pm$ 0.10  & 14.77 $\pm$ 0.20  & 15.53 $\pm$ 1.00  & 4.6    & 0.09  & 302      \\ [1ex]    
B318-G042   & ~$00^h$ $40^m$ $00.80^s$ & $+40\degr$ $34\arcmin$ $09.06\arcsec$ & 16.17 $\pm$ 1.00  & 16.39 $\pm$ 0.66  & 15.49 $\pm$ 1.00  & 3.8    & 0.29  & 1905     \\ [1ex]    
B321-G046   & ~$00^h$ $40^m$ $15.33^s$ & $+40\degr$ $27\arcmin$ $45.98\arcsec$ & 17.11 $\pm$ 0.45  & 15.88 $\pm$ 0.57  & 15.18 $\pm$ 0.29  & 4.2    & 0.13  & 200      \\ [1ex]    
B327-G053   & ~$00^h$ $40^m$ $24.12^s$ & $+40\degr$ $36\arcmin$ $22.38\arcsec$ & 14.91 $\pm$ 0.07  & 14.32 $\pm$ 0.10  & 14.14 $\pm$ 0.15  & 4.5    & 0.06  & 2754     \\ [1ex]    
B376-G309   & ~$00^h$ $45^m$ $48.38^s$ & $+41\degr$ $42\arcmin$ $39.87\arcsec$ & 16.59 $\pm$ 0.18  & 16.07 $\pm$ 0.80  & 16.02 $\pm$ 1.00  & 4.1    & 0.09  & 295      \\ [1ex]    
B448-D035   & ~$00^h$ $40^m$ $36.52^s$ & $+40\degr$ $40\arcmin$ $14.94\arcsec$ & 16.51 $\pm$ 0.34  & 16.45 $\pm$ 1.00  & 15.66 $\pm$ 1.22  & 4.1    & 0.16  & 115      \\ [1ex]    
B475-V128   & ~$00^h$ $44^m$ $55.92^s$ & $+41\degr$ $54\arcmin$ $00.33\arcsec$ & 15.10 $\pm$ 0.08  & 14.68 $\pm$ 0.12  & 14.38 $\pm$ 0.17  & 4.7    & 0.07  & 1445     \\ [1ex]    
V031        & ~$00^h$ $41^m$ $12.17^s$ & $+41\degr$ $05\arcmin$ $30.21\arcsec$ & 14.80 $\pm$ 0.06  & 14.42 $\pm$ 1.00  & 13.77 $\pm$ 0.11  & 4.8    & 0.10  & 1230     \\ [1ex]    
B083-G146   & ~$00^h$ $42^m$ $16.46^s$ & $+41\degr$ $45\arcmin$ $20.53\arcsec$ & 14.88 $\pm$ 0.05  & 14.62 $\pm$ 0.12  & 14.07 $\pm$ 0.13  & $>$4.7 & $\dots$ & $\dots$  \\ [1ex] 
B222-G277   & $00^h$ $44^m$ $25.29^s$  & $+41\degr$ $14\arcmin$ $11.62\arcsec$ & 15.27 $\pm$ 0.13  & 14.41 $\pm$ 0.09  & 14.16 $\pm$ 0.08  & $>$4.6 & $\dots$ & $\dots$  \\ [1ex] 
B347-G154   & $00^h$ $42^m$ $22.89^s$  & $+41\degr$ $54\arcmin$ $27.40\arcsec$ & 14.68 $\pm$ 0.05  & 14.17 $\pm$ 0.04  & 14.17 $\pm$ 0.18  & $>$4.7 & $\dots$ & $\dots$  \\ [1ex] 
B374-G306   & $00^h$ $45^m$ $44.53^s$  & $+41\degr$ $41\arcmin$ $55.10\arcsec$ & 17.21 $\pm$ 0.50  & 18.50 $\pm$ 0.82  & 16.32 $\pm$ 0.84  & $>$3.9 & $\dots$ & $\dots$  \\ [1ex] 
NB16        & $00^h$ $42^m$ $33.11^s$  & $+41\degr$ $20\arcmin$ $16.48\arcsec$ & 14.91 $\pm$ 0.09  & 14.11 $\pm$ 0.07  & 13.46 $\pm$ 0.11  & $>$4.8 & $\dots$ & $\dots$  \\ [1ex] 
\hline                                                                                                                                                     
P09b\\                                                                                                                                                     
\hline                                                                                                                                                     
\\                                                                                                                                                         
B049-G112   & $00^h$ $41^m$ $45.59^s$ & $+40\degr$ $49\arcmin$ $54.53\arcsec$ & 15.53 $\pm$ 0.13  & 15.27 $\pm$ 0.23  & 14.42 $\pm$ 0.06  & 4.5       & 0.09        & $\dots$  \\ [1ex]       
B367-G292   & $\dots                $ & $\dots$                               & $\dots$           &  $\dots$          &  $\dots$          & [4.3]$^a$   & [0.11]    & $\dots$  \\ [1ex]    
B458-D049   & $00^h$ $41^m$ $44.60^s$ & $+40\degr$ $51\arcmin$ $20.40\arcsec$ & 16.69 $\pm$ 0.35  & 15.04 $\pm$ 0.15  & 14.96 $\pm$ 0.15  & 4.1       & 0.15        & $\dots$  \\ [1ex]       
B521        & $00^h$ $41^m$ $41.80^s$ & $+40\degr$ $52\arcmin$ $02.41\arcsec$ & 17.32 $\pm$ 0.51  & 16.27 $\pm$ 0.43  & 16.28 $\pm$ 0.60  & 3.9       & 0.16        & $\dots$  \\ [1ex]       
M039        & $\dots                $ & $\dots$                               & $\dots$           & $\dots$           & $\dots$           & [3.8]$^a$   & [0.16]    & $\dots$  \\ [1ex]    
M050        & $00^h$ $44^m$ $40.83^s$ & $+41\degr$ $30\arcmin$ $09.68\arcsec$ & 16.14 $\pm$ 0.14  & 14.90 $\pm$ 0.19  & 15.01 $\pm$ 0.31  & 4.3       & 0.13        & $\dots$  \\ [1ex]       
\hline
WH01\\
\hline  								    
\\
B315-G038   & $00^h$ $39^m$ $48.51^s$ & $+40\degr$ $31\arcmin$ $30.33\arcsec$ & 14.99 $\pm$ 0.09  & 14.49 $\pm$ 0.10  & 14.24 $\pm$ 0.09  & 4.6       & 0.05    &  4074    \\ [1ex]   
B319-G044   & $00^h$ $40^m$ $03.03^s$ & $+40\degr$ $33\arcmin$ $58.25\arcsec$ & 16.30 $\pm$ 0.12  & 15.94 $\pm$ 0.47  & 16.78 $\pm$ 0.52  & 3.9       & 0.10    &  182    \\ [1ex]   
B342-G094   & $00^h$ $41^m$ $24.15^s$ & $+40\degr$ $36\arcmin$ $48.55\arcsec$ & 16.67 $\pm$ 0.48  & 15.57 $\pm$ 0.38  & 16.94 $\pm$ 1.00  & 4.0       & 0.17    &  214    \\ [1ex]   
B368-G293   & $00^h$ $44^m$ $47.50^s$ & $+41\degr$ $51\arcmin$ $09.39\arcsec$ & 15.89 $\pm$ 0.27  & 15.14 $\pm$ 0.35  & 14.60 $\pm$ 0.21  & 4.4       & 0.08    &  251    \\ [1ex]   
     
\hline                                                                     
\end{tabular}                                                               
                     
{\begin{flushleft}{}
\item In a few cases the data allowed us to obtain only a lower limit to the 
mass. $\alpha_{J2000}$ and $\delta_{J2000}$ are from 2MASS-6X-PSC 
catalog, J, H, K are from r=10\farcs0 ap. phot. in the 2MASS-6X-PSC 
catalog. Note that $err_{JHK}$=1.00 corresponds to  $err_{JHK}$=null in the 
2MASS-6X-PSC catalog.
\item[$^{\mathrm{a}}$] Estimated from Fig.~\ref{Mvage}, as these clusters lack NIR photometry. These mass estimates will not  be used in the following to preserve the homogeneity of the sample.
\end{flushleft}}

\end{table*}


   \begin{figure}
   \centering
   \includegraphics[width=9cm]{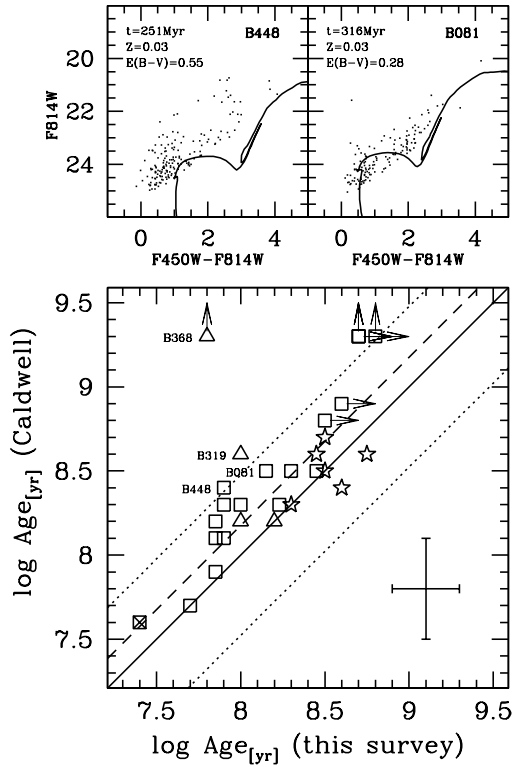}
      \caption{Bottom panel: comparison of the CMD-based ages from Tab.~2 
      with the ages obtained by C09 from integrated spectra. 
      The symbols are the same as in Fig.~\ref{Mvage}.
      B257D is not plotted because it is not included in the C09 sample.
      The error bars show the average errors.      
      The vertical arrows indicate clusters defined as ''older'' than 2 Gyr by 
      Caldwell et al. (2009). The two clusters from our own survey 
      for which the two independent
      estimates show the greatest difference are labeled (B448 and B081).
      Top panels: Comparison of the observed CMD for B448 and B081 with
      the isochrone corresponding to the age, metallicity and reddening 
      estimates provided by C09 for these clusters (values reported in the upper
      left corner of each panel). 
      Note that in the case of B448 the reddening estimated by C09 is 
      obviously too large, while in the case of B081, the 
      metallicity assumed by C09 (Z=0.03 for all the clusters) seems
      the principal responsible for the mismatch.  
      }
         \label{confr_age}
   \end{figure}
%

\subsection{Comparisons with Caldwell et al. (2009)}

A comparison of the results obtained here from the analysis of our HST-WFPC2
CMDs with those of the extensive and the independent analysis by C09, based on
high-quality integrated spectra is clearly worthwhile, in this context.

In the lower panel of Fig.~\ref{confr_age}, the age estimates from Table.~\ref{table:2}
are compared with those by C09. The two set of ages do agree within the
uncertainties, but there is a clear systematic offset as C09 ages are larger than those
listed in Tab.~\ref{table:2} by a factor of $\simeq 1.5$, in average, and up to 
a factor of $\ga 3$ in the worst case (we are considering only clusters having
age estimates in both sets, not lower limits). We note that this systematic offset
occurs also if one restricts the sample by WH01, and also to the three
clusters for which C09 provides CMD-based age estimates of their own (see their
Tab.~7), hence it is a characteristic feature of their spectroscopic age
estimates.

A difference that may produce a systematic 
offset between our ages and those by C09 is
that they adopt super-solar metallicity models ($Z=0.04$) for all the 
clusters,
while we leave metallicity as a free parameter of our fit and, in fact, we 
adopt solar or less-than-solar metallicity models in all cases (see
Tab.~\ref{table:2}). If both sets of ages were derived from isochrones
fitting the effect should be the opposite, i.e. a younger isochrone is 
required
to fit a given CMD with a model of higher metallicity. However it is not
clear if this general behavior is shared also by models of integrated spectra.

In the upper panels of Fig.~\ref{confr_age} we show the two cases (among those
included in our own survey) that display the widest difference between the two
age estimates. We superposed on the observed CMDs the isochrones corresponding to
the best-fit estimates by C09, corrected by the reddening provided by these
authors. The case of B448 shows very clearly that the solution provided by C09
significantly overestimates the reddening, and it is not compatible with the observed CMD.
In the case of B081, the comparison suggests that the choice of super-solar
metallicity models by C09 may be particularly unsuitable for this cluster,
leading to a larger-than-average error in the age estimate. 

Two cases of
especially remarkable differences occur also with the set by WH01 (open 
triangles in Fig.~\ref{confr_age}). B319=G44 is considered also in Tab.~7 of C09,
where a spectroscopic age of 0.28 Gyr is reported, to be compared to the 
CMD-based age estimated of 0.10 Gyr by WH01. Moreover the reported 
spectroscopic value is most probably a typo, as in Table~2 of C09 
(their primary source of cluster ages) they report log(age)=8.6 for B319=G44, 
corresponding to 0.398 Gyr (the value that is plotted in Fig.~\ref{confr_age}).
In any case, the spectrum appears to be reasonably fitted by a 
Z=0.04, age=500 Myr model (N. Caldwell, private communication), while the 
CMD shown by WH01 is clearly not compatible with such an old age. The a-priori assumption of 
super-solar metallicity models by C09 may  also  be the origin of this mismatch.
The case of B368=G293 (not included in Tab.~7 of C09), that is classified by 
C09 as ''older than 2 Gyr'' while the CMD by WH01 indicates age $\la 80$ Myr, has to be ascribed to a typographical error by C09; in fact the cluster was not observed by that authors (N. Caldwell, private communication).

\begin{figure}
   \centering
   \includegraphics[width=9cm]{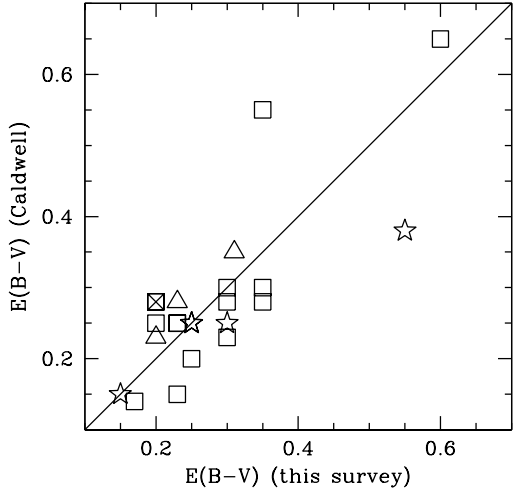}
      \caption{Comparison of the E(B-V) estimates from Tab.~3 with those by
      C09. The symbols are the same as in Fig.~\ref{Mvage}.}
         \label{confr_ebv}
   \end{figure}

\begin{figure}
   \centering
   \includegraphics[width=9cm]{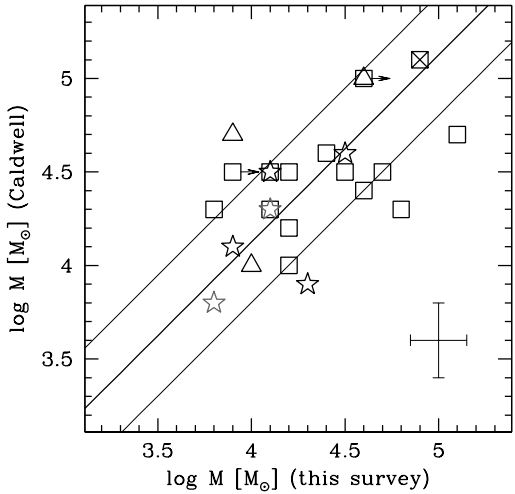}
      \caption{Comparison of the masses estimates from Tab.~3 with 
      those by C09. The symbols are the same as in Fig.~\ref{Mvage}.
      The grey symbols show the clusters that have "null" error 
      on IR magnitudes in the 2MASS-6X-PSC catalog.
      The thick line is the $M_{t.s.}=M_{C09}$ locus, the thin lines
      bracket the $\pm 1 \sigma$ range about this locus.
       The error bars show the average errors.}
	 \label{confr_mass}
   \end{figure}

Fig.~\ref{confr_ebv} shows the comparison between our estimates of E(B-V) and
those by C09. In this case as well there is reasonable overall agreement, most of
the differences being within the uncertainties. The most discrepant case
is B448, already discussed above (see Fig.~\ref{confr_age}). Finally, in
Fig.~\ref{confr_mass} the mass estimates are compared. Also in these cases the
two set of estimates agree within the uncertainties (1 $\sigma$ is a factor of
2.4), the strongest discrepancy is to be attributed to the overestimate of the age
for B319=G44 by C09 discussed above.

In conclusion, while we are unable to identify the reason of the (modest)
systematic overestimate of the ages by C09, it has to be concluded that the
agreement between the two independent sets of age, reddening, and mass estimates
is quite satisfactory, if the observational uncertainties are taken into the due account.


%

%

\section{Summary and Discussion}
\label{disc}

We presented the main results of a survey aimed at the determination of the nature  of a sample of 20 candidate YMC in the thin disk of M31 
(one of which, VdB0, was studied in Pap-I).
One of the targets surveyed  turned out to be a bright star 
projected onto the dense disk of M31, and thus erroneously 
classified as a possible cluster. All the other targets were revealed to be 
genuine star clusters and we were able to obtain  reliable CMDs for all of them. The main results from our own survey can be summarized as follows:

\begin{enumerate}

\item New integrated-light spectroscopy became available for many of our targets since the original selection was performed. Three of them (B083, NB16 and B347) were revealed by the new data to be not good YMC candidates as defined by F05. The CMDs obtained in this study confirms that they are likely old clusters.

\item Among the remaining 17 targets, 16 are genuine clusters and one is in fact a star (NB67), as said above. Thus the fraction of spurious objects in our well-defined sample of BLCC=YMC is just 1/16 = 6.2\%. Even excluding the two clusters considered at point 3., below, the incidence remains below 10\%. 
The extended sample considered in Appendix~\ref{appB} fully confirms these  results.
We must conclude that M31 YMC are not especially plagued by contamination from spurious sources and most of the clusters considered in the original analysis by F05 should be real\footnote{It may be useful to stress again that the clusters of our survey were selected among the class f=1 RBC entries, see Sect.~\ref{sample} and Galleti et al.~(\cite{svr}).}. In particular, {\em asterisms}, suggested as a possible major contaminant of the sample by C06, are in fact found to be not a particular reason of concern, in this context (see also the discussion by C09). 

\item Two of the sixteen genuine clusters (B374 and B222) have 
integrated properties compatible with being YMCs but they do not show 
a detectable MS in the range of magnitudes sampled by our CMDs. We can 
provide only an upper limit to the age of these clusters ($\ga 300$ Myr), 
but the available data suggest that they are good candidate 
intermediate-age clusters that indeed would merit follow-up with deeper HST photometry.

\item The fourteen confirmed young clusters (including VdB0, studied in Pap-I) 
show a clear MS in the range of magnitudes sampled by our CMDs, 
hence we were able to obtain reliable estimates of their ages, reddenings and 
(an educated guess of) metallicities by comparison of the observed CMD and LF 
with theoretical models.
Ten of them have ages in the range 25-100 Myr, the other four range 
between 140 Myr and 280 Myr. The adopted metallicities include 
$Z=0.004$ (one case), $Z=0.008$ (three cases), and $Z=0.019$ 
(solar metallicity, ten cases). The estimated reddenings range from E(B-V)=0.06 to E(B-V)=0.60, with E(B-V)=0.20-0.30 as most typical values.

\end{enumerate}


   \begin{figure}
   \centering
   \includegraphics[width=9cm]{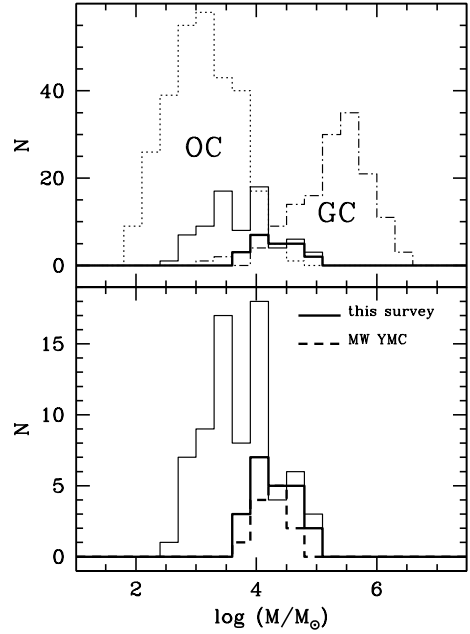}
      \caption{Upper panel: The mass distribution of the sample of YMC 
      studied here (from Tab.~\ref{table:3}, thick continuous
     line) is compared with the mass distribution of Galactic OCs
      (dotted line) and Galactic globular clusters (dashed lines). Masses of
      Galactic clusters are from B08. 
      Lower panel: zoomed view of the 
      distribution   of M31 YMC  compared with the distribution of the YMC of 
      the Milky Way (data from M09).}
         \label{histo}
   \end{figure}
%


   \begin{figure*}
   \centering
   \includegraphics[width=16cm]{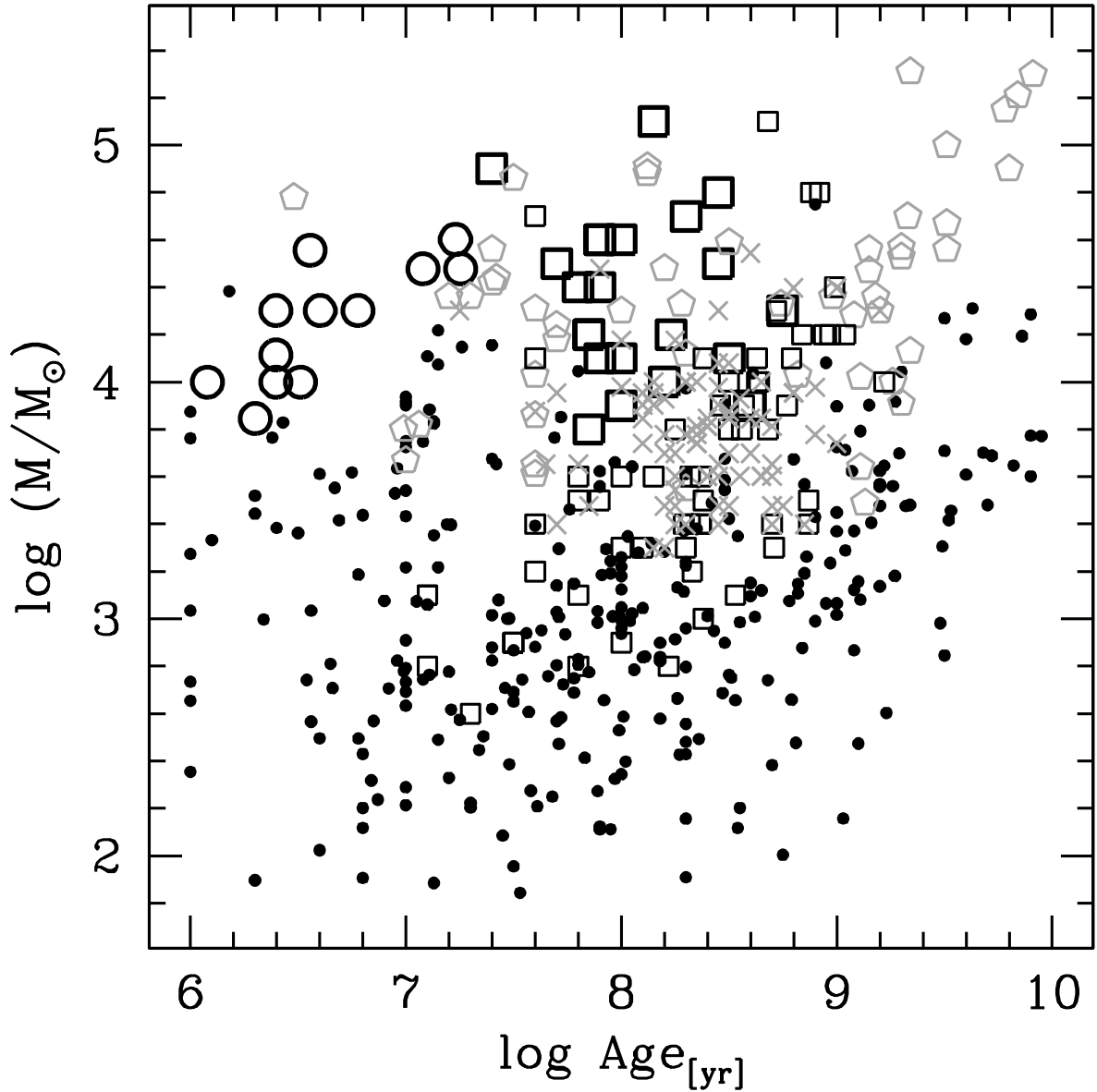}
      \caption{Comparison between Galactic OCs (small filled circles), M31 YMC from 
      the present study (big open squares), MW YMC from 
      M09 (big open circles), M31's clusters from pap-II (small open squares), 
      Magellanic Clouds clusters (grey open pentagons), and M33's clusters (grey crosses) in the 
      log(age) $vs.$ log Mass plane. 
      Masses of Galactic OCs are from B08, masses of Magellanic Clouds clusters are from \cite{mclaug}
      and masses of M33 clusters are from \cite{sanrom}. For M33 and the 
      Magellanic Clouds  only clusters younger then 10 Gyr are shown.} 
         \label{agemass}
   \end{figure*}
%

To increment our final sample of YMC we included ten further clusters for which the age was estimated from their CMDs (obtained from HST imaging) with methods  strictly homogeneous with those adopted here, from WH01 and P09b. In this way we assembled a final sample of 24 confirmed young clusters. For 22 of these we were able to obtain 
reliable estimates 
of the total stellar mass by coupling our age estimates with the 
integrated J,H,K magnitudes taken from the 2MASS-6X catalog.
These clusters have masses ranging from $0.6\times
10^4 M_{\sun}$ to $6\times 10^4 M_{\sun}$, with an average of 
$\sim 3\times 10^4 M_{\sun}$\footnote{The remaining two clusters, that lack NIR photometry, also have masses lying in the same range, according to the estimates obtained using the integrated V magnitude instead of J,H,K ones.}. Our estimates of ages and masses are in good
agreement with recent independent studies based on integrated light spectra (see also Pap-III for the comparison with the results by Pfalzner \cite{pfalz}).
 
\subsection{The nature of M31 YMC}

In the upper panel of Fig.~\ref{histo} the mass distribution of our 
extended sample of M31 YMCs is compared with the distributions of Galactic 
OCs and GCs (masses from B08). The clusters considered here appear to 
lie in the middle of the two distributions, overlapping with the high-mass 
end of the OCs and with the low-mass end of GCs. This comparison 
provide a further confirmation that the YMCs (=BLCCs) of M31 are indeed 
more similar to the YMCs of the LMC than to classical 
OCs of the Milky Way, i.e. the original hypothesis advanced in F05. 
This is in full agreement with the main conclusions by C09, obtained 
with a completely independent method (less sensitive to age than ours) on a wider sample.

The lower panel of Fig.~\ref{histo} compares our clusters with the YMCs seen toward the center of the Milky Way as listed by M09. The two samples have very similar mass distributions, suggesting that they are also similar in nature. 
An obvious difference between the two sets of clusters was already suggested 
in Pap-I and is confirmed here: the M31 YMCs of our sample are significantly 
older that the YMC discovered until now in the Galaxy ($\ga 50$ Myr vs. $\la 20$ Myr; see below for possible explanations).
We confirm that the M31 YMCs studied here have larger sizes (half-light-radii) with  respect to their MW counterparts (see Pap-I and Pap-III); this seems in agreement with the age-size relations proposed by Pfalzner~(\cite{pfalz}; see Pap-III for discussion).

A more thorough comparison between various samples of YMCs is presented in  
Fig.~\ref{agemass}, where Galactic OCs and YMCs, YMCs from M33 
(San Roman et al.~\cite{sanrom}; for further discussion on M33's star clusters
see Sarajedini \& Mancone \cite{sarajed}, Zloczewski et al.~\cite{zlocz},
Park et al.~\cite{park}), the LMC, the Small Magellanic Cloud 
(McLaughlin \& van der Marel \cite{mclaug}), and  M31 are plotted 
together in a log(age) $vs.$ log Mass diagram. Fig.~\ref{agemass} is 
affected by a number of selection effects that deserve to be described in some detail.  

\begin{enumerate}

\item The minimum mass threshold appears to increase with age (at least for
age $\ga$ 10 Myr, see the Galactic OCs if Fig.~\ref{agemass}): this is 
due to the fact that the lower the mass of a cluster, the shorter is its dissolution time, as the cluster is less resilient to all the internal and external effects that may lead to its disruption (Gieles et al. \cite{gieles}, Pap-III, and references therein). The minimum mass threshold  for samples in external galaxies is obviously due to the inherent magnitude limits. 

\item Also the maximum mass threshold increases with age in log~Age vs. log~Mass plots (Hunter et al. \cite{hunter}; Gieles \cite{giel09}; the effect is clearly evident in Fig.~\ref{agemass} if one looks at the MW OCs, that cover the widest range in ages). This general behavior can be easily explained as a simple consequence of varying the sample size as a function of the age bin in the logarithmic scale.
Assuming a power-law mass function and a constant Cluster Formation Rate (CFR) the number of cluster per logarithmic age bin increases with age. For an exponent of the power law mass function ($N(M)\propto M^{-\alpha}$) 
$\alpha=2$, that is a reasonable approximation for most of the observed cluster systems, log~$M_{max} \propto$ log~Age (see Gieles \cite{giel09}, for detailed discussion and references).

\item While the lack of massive ($M\ga 10^4 M_{\sun}$) clusters older than 
400 Myr in the Milky Way is probably real, the typical limiting magnitude 
($V\sim 27$, Rich et al. \cite{rich}) of available CMDs of M31 clusters 
prevent us from drawing firm general conclusions about objects in that age range in M31. The cases of B222 and B374, treated here, are excellent examples of clusters that may populate that region of the diagram but lack a reliable age estimate because the available photometry is too shallow (see Puzia et al.~\cite{puzia}).

\item The lack of massive (log~$(M/M_{\sun})>3.6$) M31 clusters younger 
than 25-50 Myr may be due to the contribution of several biases. First, such young clusters may be hard to select from the RBC as there are no objects bluer than $(B-V)_0\simeq 0.0$  in the list of confirmed clusters (see F05). This is not surprising as the RBC was intended to be a catalog of globular clusters.
Second, for ages $\la$ 8 Myr the $H_{\beta}$ index is expected to fall below the threshold adopted to select YMC candidates (see, for example, Fig.~7 of F05), thus (possibly) preventing the selection of these objects for our survey.
Third, very young objects should have their luminosity dominated by a few 
massive stars near their centers, thus leading to objects that may appear 
more like 
blended stars than like a star cluster at the distance of M31, even in HST images, thus preventing their inclusions in lists of candidate YMCs.
Fourth, it can be hypothesized a positive correlation between the age of the clusters and their height above the disk plane, such that the youngest clusters are more deeply embedded in the thin dust layer of the M31 disc, out of our reach even from our privileged point of view, while most/some of the older clusters would be visible just because they lie above the densest part of that layer.
There are indications that this kind of correlation actually holds in our own Galaxy (V.D. Ivanov, private communication).

\item The lack of massive (log~$(M/M_{\sun})>3.6$) MW clusters {\em older} than 25-50 Myr may also be associated with an observational bias. Galactic YMC have been identified as clumps of bright stars in the near and mid IR and the youngest clusters, having the brightest RSG, are easier to detect in this way. Moreover the sample of Open/YM Galactic clusters is limited (essentially by the effect of interstellar extinction in the Galactic disc) to a volume of a few kpc around the Sun, while M31 (or M33) YMCs can be selected over the whole disk of their parent galaxy, thus introducing a bias that favors the detection of rarer cluster species 
(massive clusters) in the latter galaxies with respect to the MW. 

\item There seems to be a significantly under-dense region in Fig.~\ref{agemass}, for masses $\ga 10^3~M_{\sun}$ and ages between $\sim 15$ Myr and 
$\sim 50$ Myr ($7.2\la$~log~Age$\la 7.7$). The same feature was noted by Whitmore et al. (\cite{whitmore}) in their study of the cluster system of the Antennae and it was attributed by a degeneracy in age dating from broad band colors occurring in that age range due to the prompt onset of the RSG phase (see 
Whitmore et al. \cite{whitmore}, for details, discussion and further references). Virtually all the clusters plotted in Fig.~\ref{agemass} had their ages estimated from the CMD of their stars (instead of broad-band colors, see also Pap-II), hence our sample should not be affected by this bias, at least in principle. However the coincidence of the feature with that noted by Whitmore et al. (\cite{whitmore}) suggests that the same kind of bias against ages in that interval may be at work also in Fig.~\ref{agemass}.

\item The samples of clusters from all the galaxies involved in Fig.~\ref{agemass} have been selected according to different criteria, by color, magnitude, etc. 

\end{enumerate}

Given all the above considerations, it does not seem possible to draw any firm conclusion from the comparison shown in Fig.~\ref{agemass}. The only straightforward conclusion is that {\em YMCs in the age range 50-500 Myr are relatively common in all the most massive star-forming galaxies of the Local Group} (M31, M33, LMC and SMC). The only exception (the Milky Way) may be ascribable to observational biases, but it 
cannot be excluded that it is instead (at least partly) associated with intrinsic properties of the Milky Way, that appears peculiar under several aspects with respect to the typical spiral galaxies (and to M31, in particular see Hammer et al.~\cite{hammer}, and  Yin et al.~\cite{yin}).
As the samples of M33 and M31 should be subject to the same kind of biases (as the distances are similar and the data have been collected with HST in both cases), the difference in the maximum mass limit between the two samples is likely real, and it can probably be ascribed to the difference in total mass between the discs of the two galaxies: larger discs should host more numerous populations of clusters, thus enhancing the probability of producing clusters with higher (maximum) masses (see Gieles \cite{giel09}, and references therein).


\subsection{Radial trends}

Given the wealth of data collected for our target clusters, it may be  
useful to look for correlations between their physical parameters, including 
their position within the M31 disc. Limiting the analysis to the young  
clusters (age $<$ 1 Gyr), that constitute a more homogeneous sample of bona-fide  
thin disk objects, it turns out that our sample is still too sparse for a 
thorough analysis of these correlations. In particular the covered ranges of  
age, mass and position are quite limited, thus not allowing us to reveal large  
scale trends, in most cases. Moreover, the adopted approach of CMD analysis 
provides just an educated guess of the metallicity of the clusters, aimed at
obtaining the most reliable estimate of the clusters age, which was the main 
objective of our analysis. These limitations prevent the possibility of a
meaningful study of the radial metallicity gradient with our data. It  
should also be recalled that the correlations bewteen the structural  
parameters of the clusters (mass, radius, density etc.) have already been 
discussed in Pap-III, hence here we consider only age, mass, de-projected 
galactocentric distance ($R_d$; assuming and inclination of $i=12.5\degr$ of 
the disk with respect to the plane of the sky, see Simien 
et al. \cite{simien} and Pritchet \& van den Bergh \cite{prit}), X, Y, and 
reddening.

\begin{figure}
   \centering
   \includegraphics[width=9cm]{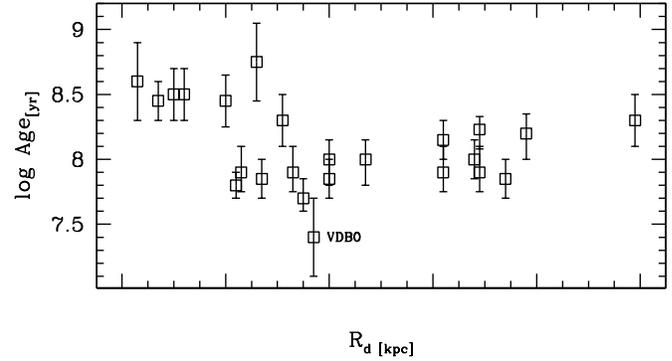}
      \caption{Age as a function of the deprojected galactocentric distance 
      for the young clusters (open squares with error bars). 
      The cluster VdB0 has been labeled as it is by far, the youngest 
      of the whole sample.}
         \label{radtrend}
   \end{figure}

Having checked all the combination of parameters, the only  
correlation  that appeared remarkable to us is presented 
in Fig.~\ref{radtrend}.  It is a trend of decreasing
age with galactocentric distance, that seems statistically significant  
if one consider the associated errors. Given the relatively limited range of
galactocentric distance covered, in our view the observed distribution  
can be interpreted in two ways:

\begin{itemize}

\item as a part of a larger trend resulting from a inside-out wave of  
cluster formation. In this case the trend toward older mean ages should  
continue at lower radii and Fig.~\ref{radtrend} shows the transition between 
a regime of decreasing age with galactocentric distance and an asymptotic 
regime of constant age in the outermost fringes of the disc;

\item more likely, as a sharp transition in the epoch of the highest  
rate of star/cluster formation occurring at the onset of the $R_d\sim 10$ kpc  
``ring of fire''. This would be consistent with the well known burst of recent  
star formation that characterize this prominent structure of the M31 disc.

\end{itemize}

While not especially conclusive or insightful, the result shown in  
Fig.~\ref{radtrend} gives a clear idea of how useful YMCs can be as tracers of 
the structure and evolution of the disk itself, in particular if large and 
reliable samples can be assembled.


%

\subsection{Final remarks}

This research has demonstrated that the conspicuous population of bright disk 
objects studied by F05 consists of genuine YMC, similar to those found in the LMC, SMC and M33 galaxies. These clusters may open a new window to the study of the recent star formation history in the disk of M31. A systematic analysis over the whole extent of the M31 disk may provide the opportunity to study a rich system of young clusters using a sample much less affected by selection biases than in our own Galaxy, and to better constrain the models of dynamical evolution of clusters within the discs of spiral galaxies. M31 YMCs like those studied here provide also an excellent tracer of the disk kinematics in that galaxy, independent of (and in addition to) the HI gas. 
Recent wide-field surveys (Vansevicius et al.~\cite{vanse}; see also Pap-II) suggest that a rich harvest of genuine YMCs await to be discovered in the disk of our next neighbor giant galaxy in Andromeda.


\begin{acknowledgements}
We are grateful to an anonymous referee for a constructive report and for
useful suggestions that improved the quality of this paper.
S.P. and M.B. acknowledge the financial support of INAF through the 
PRIN 2007 grant CRA 1.06.10.04 ``The local route to galaxy
formation...''. 
P.B. acknowledges research support through a Discovery Grant from the Natural 
Sciences and Engineering Research Council of Canada.
J.G.C. is grateful for partial support through
grant HST-GO-10818.01-A from the STcI.
T.H.P. gratefully acknowledges support in  
form of a Plaskett Fellowship at the Herzberg institute of  
Astrophysics in Victoria, BC.
J.S. was supported by  NASA through an Hubble Fellowship, 
administered by STScI. We are grateful to S. van den Bergh for having pointed 
out some errors in the historical reconstruction of the discovery of VdB0 that 
were  reported in a previous version of the paper.
We are grateful to M. Gieles, V.D. Ivanov, N. Caldwell, and, in particular, to M. Messineo for useful discussions and suggestions.
\end{acknowledgements}

\clearpage

\begin{appendix}
\section{RBC clusters serendipitously imaged in our survey}
\label{appe}

To ascertain the real nature of candidate M31 clusters proposed by various authors is a daunting but necessary task to keep cluster catalogs as complete and clean as possible from spurious sources. There are several criteria that may be used to check candidates (see Galleti et al.~\cite{svr} for references and discussion), but resolving them into stars by means of high spatial resolution imaging is by far the safest method of all. In addition to the clusters that were the main target of our survey, and to the low-luminosity clusters identified by 
Hodge et al.~\cite{lessclu}, our WFPC2 images serendipitously 
included several clusters and candidate clusters listed in the RBC. 
Inspection of our images allowed us to place their classification on  
firmer footing. The results of this analysis are summarized in Table~\ref{tab:app}. Their classification in the RBC has been modified accordingly.
In Table~\ref{tab:app} we report the name of the  object (column 1, name), 
the classification flag originally reported in the RBC (col.~2, f), 
the name of the cluster that was the original target of the images 
(col.~3, field), a flag indicating if the object was imaged with the PC or with one of the WF cameras (col.~4, chip), and, finally, a comment on its classification as derived from the inspection of the new images. In some case the classification 
remains uncertain (comments with ``?''). In some cases the image reveals that the object is extended but do not clarify its nature (cluster/galaxy/HII region etc.), in these cases we report the comment ``not a star''. An estimate of the radial velocity will suffice to definitely establish if these objects are M31 clusters or background galaxies (see Galleti et al.~\cite{svr}).
In some cases, some clusters that were among the main targets of our survey were serendipitously re-imaged in the WF field surrounding other targets. For obvious reasons these cases are not reported in Table~\ref{tab:app}. 
On the other hand some clusters have been serendipitously imaged in two different pointings: in these cases we report the classification derived from both sets of images. 
Some of the clusters of Table~\ref{tab:app} were independently re-identified in Pap-II (B061D, B319, B014D, B256D, DAO84), for two of them a meaningful CMD was also 
obtained there (B061D and B319); this lends additional support to the reliability of their classification.
Finally, we reported in the table also some clusters whose nature was already confirmed by previous HST imaging, for completeness (see the case of B319=G044, observed by WH01).

It may be interesting to note that among the 19 RBC class f=2 
(candidate clusters) objects listed in Tab.~\ref{tab:app}, 3 turn out 
to be real clusters (or likely clusters), 5 are extended objects that 
lack the $v_r$ measure needed to ultimately establish their membership to M31, while 11 are non-clusters 
(or likely non-clusters), most of them being stars. According to this limited sample it can be concluded that the fraction of genuine M31 clusters among class f=2 entries of the RBC ranges from 
$\frac{3}{19}$=16\%$\pm$ 14\% to $\frac{8}{19}$=42\%$\pm$ 12\%. These numbers 
should be considered as 
somewhat pessimistic as they are computed on a sample of clusters projected on 
the densest regions of the M31 disc, where the probability of contamination  
from bright stars of M31 is at its maximum. To give a rough idea of the 
number of genuine clusters that are still hidden among the candidates listed  
in the RBC  one can take the 16\% of the number of class=2 RBC entries, 
i.e. $0.16\times 1049\simeq 168$. A significant fraction of these may be 
YMCs ($\ga$ 15\%, according to F05).

Considering the objects listed in Tab.~\ref{table:1} and Tab.~\ref{tab:app}, the survey images allowed us to verify the nature of 25 objects classified as genuine clusters (class f=1) in the RBC. We confirm that 23 of them are real clusters while 2 are (one or two) stars. From this number one can estimate the fraction of spurious sources among class f=1 RBC entries as $\frac{2}{25}$=8\%$\pm$8\%, that is remarkably low and is in excellent 
agreement with the estimate by G09 that finds $\la$4\% from a sample of 252 objects.

Considering the fraction of real clusters among class f=1 entries as 92\% and that among f=2 entries as 16\%, the expected number of genuine M31 clusters in the RBC (GC+YMC) is estimated as $\sim 630$, while the number of old clusters (GCs) should be $\sim 530$, in reasonable agreement with the results by Barmby et al.~\cite{barm00} and F05. Note that, at present, the number of confirmed (likely) old
clusters (f=1 and y=0) in the RBC is 418; correcting this for contamination leads to 384 bona-fide GCs, more than double than the number of GCs encountered in the Milky Way galaxy ($\simeq 150$, Harris~\cite{h96}).  

\begin{table*} 
\centering
\caption{RBC clusters serendipitously imaged in our survey.}
\label{tab:app}
  \begin{tabular}  {@{}lcccc@{}}
  \hline \hline
Name  &  f$^1$ & Field  & Chip  & Comment \\ 
 \hline
 \hline 
 \\   
  B014D  & 2 & B015D  & PC &    cluster        \\ [1ex]  
  B061D  & 2 & NB16   & WF &    cluster        \\ [1ex]  
  B256D  & 2 & B257D  & WF &    cluster$^2$       \\ [1ex]  
  B256D  & 2 & B475   & WF &    cluster$^2$       \\ [1ex]  
 SK067B  & 2 & B015D  & WF &    not a star     \\ [1ex]  
 SK071C  & 2 & B475   & WF &    not a star     \\ [1ex]  
 SK185B  & 2 & B475   & WF &    not a star     \\ [1ex]  
  B068D  & 2 & NB16   & WF &    not a star     \\ [1ex]  
  B068D  & 2 & NB67   & WF &    not a star     \\ [1ex]  
  B019D  & 2 & V031   & WF &    not a star     \\ [1ex]  
   NB64  & 2 & NB16   & WF &    star?	  \\ [1ex]  
   NB64  & 2 & NB67   & WF &    star?	  \\ [1ex]  
 SK091B  & 2 & B066   & WF &    star	  \\ [1ex]  
  B048D  & 2 & B081   & PC &    star	  \\ [1ex]  
 SK091C  & 2 & B374   & WF &    star	  \\ [1ex]  
 SK188B  & 2 & B475   & WF &    star	  \\ [1ex]  
   NB47  & 2 & NB16   & WF &    star	  \\ [1ex]  
 SK083B  & 2 & B043   & WF &    2 stars + nebula?     \\ [1ex]
  B057D  & 2 & NB16   & WF &    2 stars     \\ [1ex]  
   NB43  & 2 & NB67   & WF &    2 stars     \\ [1ex]  
  B192D  & 2 & B327   & WF &    galaxy      \\ [1ex]  
 SK194C  & 2 & B376   & WF &    galaxy      \\ [1ex]  
   B376  & 1 & B374   & WF &    cluster 	\\ [1ex]  
  B257D  & 1 & B475   & WF &    cluster 	\\ [1ex]  
   B319  & 1 & B318   & WF &    cluster     \\ [1ex]  
  DAO84  & 1 & B374   & WF &    not a star$^3$	       \\ [1ex]  
  DAO84  & 1 & B376   & WF &    not a star$^3$	\\ [1ex]  
 SK047A  & 1 & B081   & WF &    two stars	\\ [1ex]  
   NB68  & 6 & NB16   & WF &    star?		\\ [1ex]  
   NB68  & 6 & NB67   & WF &    star?	    \\ [1ex]  
   B113  & 6 & NB16   & WF &    star?	    \\ [1ex]
 SK069D  & 6 & B083   & WF &    star	    \\ [1ex]  
  B185D  & 6 & B318   & PC &    star	    \\ [1ex]  
 SK046D  & 6 & B327   & WF &    star	    \\ [1ex]  
  B065D  & 6 & NB67   & WF &    star	    \\ [1ex]  
 SK041D  & 6 & B321   & WF &    two stars     \\ [1ex]  
   B121  & 3 & NB16   & WF &    star?	    \\ [1ex]  
   B121  & 3 & NB67   & WF &    star	    \\ [1ex]  
 \hline
\end{tabular}                                                             
\begin{list}{}{}
\item $^1$ f is the original RBC classification flag (1 globular cluster, 2 
candidate globular cluster, 3 controversial object, 6 star/s). 
\item $^2$ While the visual inspection of the images does not permit a clear cut classification, the objective analysis performed in Pap-II recognizes B256D as a star cluster.
\item $^3$ DAO84 has a radial velocity estimate that clearly identifies it as a member of M31 (see the RBC). 
\end{list}
\end{table*}

\section{Other candidate M31 YMCs with archival HST imaging}
\label{appB}

Before selecting the actual targets for our survey we searched the HST archive for YMC candidates, as listed in Tab.~1 (or Tab.~2) of F05, that had already been (serendipitously) imaged from HST. As the nature of these objects (cluster / asterism / star) can be determined from existing images  they were not included in our final list of targets. In Tab.~B.1. (referring to objectively selected candidates from Tab.~1 of F05) and 
Tab.~B.2. (referring to candidates suggested from various authors adopting 
different criteria, from Tab.~2 of F05) we list the results of that research. 
In these tables we report (1) the cluster name(s), (2) the HST program 
number(s) of the retrieved images, (3) the instrument(s) and (4) the 
filter(s) used to obtain the inspected images, (5) the classification of 
the object based on the inspection of the HST images, following the approach 
adopted in Tab.~\ref{tab:app}, above, and, finally, (6) the classification 
provided by C09 based on their spectra and/or on ground-based 
imaging (S indicates that the objects was classified by from its 
spectrum, I indicates that the object was classified with imaging, SI means 
that both imaging and spectrum were considered for the 
classification, according to C09).
At the epoch when the table was compiled (September 2009), 36 out of the 66 objects listed in Tab.~1 of F05 (including those studied in this paper) had one (or more) images in the HST archive: 34 of them are recognized as {\em real} star clusters from the inspection of the available HST images, while 2 are stars. This leads to a fraction of spurious objects in the sample of 5.5\% $\pm$ 4.0\%, in full agreement with the fraction we obtained from our original sample (Sect.~\ref{sample}). Analogously, 14 out of 21 objects listed in Tab.~2 of F05 (including those studied in this paper) had one (or more) image(s) in the HST archive: 13 of them are recognized as {\em real} star clusters from the inspection of the available HST images, while 1 is a star. This leads to a fraction of spurious objects in the sample of 7.1\% $\pm$ 7.4\%, again in full agreement with the fraction we obtained from our original sample (Sect.~\ref{sample}) and with the above results. Note that (a) all the classifications we obtained from HST imaging confirm those
independently obtained by C09 for the same objects, and (b) all the objects 
listed in Tab.~B.2. were classified as clusters by some other 
author before (see F05).

Of the 37 objects in Tab.~B.1. and Tab.~B.2. lacking HST-based classification, 31 are classified as {\em clusters} by C09; the remaining 6 have uncertain classification. Coupling the results from HST and C09 it turns out that 60 of the 66 objects from Tab.~1 of F05 are real clusters, two are stars, and four have uncertain classification; 18 of the 21 objects from Tab.~2 of F05 are real clusters, 
one is a star, and two have uncertain classification. We thus conclude that
the large majority ($\ga$90\%) of the objects identified (or proposed) by F05 as (possibly) young clusters are indeed genuine star clusters.
Finally, three clusters listed in the RBC but not comprised in the study by 
F05 where found in Pap-II to have age $< 1$ Gyr (B014D, B061D, B256D).

\scriptsize
\begin{longtable} {@{}lccccc@{}}
\caption{Classification of candidate young clusters listed in Tab.~1 of F05.}\\ 
\label{tab:app2}
\kill
\hline \hline
\\
Name  &    Obs-ID & Camera  & Filters  & Class HST & Class C09\\
\\      
\hline 
\endfirsthead
\caption{continued.}\\
\hline\hline
\\
Name  &    Obs-ID & Camera  & Filters  & Class HST & Class C09\\
\\
\hline
\\
\endhead
\hline
\endfoot
%
\\
 B008-G060      & 10407             & ACS/WFC  &F606W F435W                    &cluster  &cluster(SI)	  \\ [1ex]
 B028-G088      &                   &          &                               &	 &cluster(SI)	  \\ [1ex]
 B040-G102      & 10818             & WFPC2    &F450W F814W                    &cluster  &cluster(SI)	  \\ [1ex]
 B043-G106      & 10818             & WFPC2    &F450W F814W                    &cluster  &cluster(SI)	  \\ [1ex]
 B047-G111      &                   &          &                               &	 &cluster(S) 	  \\ [1ex]
 B049-G112      & 10407(10631)      & ACS/WFC  &F435W F606W                    &cluster  &cluster(SI)     \\ [1ex]
 B057-G118      & 10407(10631)      & ACS/WFC  &F435W F606W                    &cluster  &cluster(SI)     \\ [1ex]
 B066-G128      &                   &          &                               &cluster  &cluster(SI)	  \\ [1ex]
 B069-G132      & 10273             & ACS/WFC  &F555W F814W                    &cluster  &cluster(SI)	  \\ [1ex]
 B074-G135      &                   &          &                               &	 &cluster(S) 	  \\ [1ex]
 B081-G142      & 10818             & WFPC2    &F450W F814W                    &cluster  &cluster(SI)	  \\ [1ex]
 B083-G146      & 10818             & WFPC2    &F450W F814W                    &cluster  &cluster(S)	  \\ [1ex]
 B091-G151      & 10273             & ACS/WFC  &F555W F814W                    &cluster  &cluster(SI)	  \\ [1ex]
 B114-G175      &  5907             & WFPC2    &F555W F814W                    &cluster  &cluster(SI)	  \\ [1ex]
 B160-G214      &  9480(10273,7426) & ACS/WFC, WFPC2 &F775W F555W F814W F606W  &cluster  &cluster(SI)     \\ [1ex]
 B170-G221      &                   &          &                               &	 &cluster(SI)	  \\ [1ex]
 B210-M11       &  9709             & WFPC2    &F606W	                       &cluster  &cluster(SI)	  \\ [1ex]
 B216-G267      &                   &          &                               &	 &cluster(SI)	  \\ [1ex]
 B222-G277      & 10818             & WFPC2    &F450W F814W                    &cluster  &cluster(SI)	  \\ [1ex]
 B223-G278      &                   &          &                               &	 &cluster(SI)	  \\ [1ex]
 B237-G299      &                   &          &                               &	 &cluster(SI)	  \\ [1ex]
 B281-G288      &                   &          &                               &	 &cluster(SI)	  \\ [1ex]
 B295-G014      &                   &          &                               &	 &cluster(S) 	  \\ [1ex]
 B303-G026      &                   &          &                               &	 &cluster(SI)	  \\ [1ex]
 B307-G030      &                   &          &                               &	 &cluster(SI)	  \\ [1ex]
 B314-G037      &                   &          &                               &	 &cluster(SI)	  \\ [1ex]
 B315-G038      &  8296             & WFPC2    & F336W F439W F555W             &cluster  &cluster(SI)	  \\ [1ex]
 B318-G042      &  8296(10818)      & WFPC2    & F336W F439W F450W F555W F814W &cluster  &cluster(SI)     \\ [1ex]
 B319-G044      &  8296             & WFPC2    & F336W F439W F450W F555W F814W &cluster  &cluster(SI) 	  \\ [1ex]
 B321-G046      & 10818             & WFPC2    & F450W F814W                   &cluster  &cluster(SI)	   \\ [1ex]
 B322-G049      &                   &          &                               &	 &cluster(SI)	   \\ [1ex]
 B327-G053      & 10818             & WFPC2    & F450W F814W                   &cluster  &cluster(SI)	   \\ [1ex]
 B331-G057      &  6699             & WFPC2    & F555W F814W                   &cluster  &cluster(SI)	   \\ [1ex]
 B342-G094      &  8296             & WFPC2    & F336W F439W F555W             &cluster  &cluster(SI)	   \\ [1ex]
 B354-G186      &                   &          &                               &	 &cluster(S) 	   \\ [1ex]
      B355      &                   &          &                               &	 &possible star(S) \\ [1ex]
 B358-G219      &                   &          &                               &	 &candidate	   \\ [1ex]
 B367-G292      & 10407             & ACS/WFC  & F435W F606W                   &cluster  &cluster(SI)	   \\ [1ex]
 B368-G293      &  8296             & WFPC2    & F336W F439W F555W             &cluster  &cluster(I)	   \\ [1ex]
 B374-G306      & 10818             & WFPC2    & F450W F814W                   &cluster  &cluster(SI)	   \\ [1ex]
 B376-G309      & 10818             & WFPC2    & F450W F814W                   &cluster  &cluster(SI)	   \\ [1ex]
 B380-G313      &                   &          &                               &	 &cluster(SI)	   \\ [1ex]
 B431-G027      &                   &          &                               &	 &cluster(SI)	   \\ [1ex]
 B443-D034      &                   &          &                               &	 &cluster(SI)	   \\ [1ex]
 B448-D035      & 10818             & WFPC2    & F450W F814W                   &cluster  &cluster(SI)	   \\ [1ex]		    
      B451      &                   &          &                               &	 &possible star(I) \\ [1ex]		    
 B453-D042      &                   &          &                               &	 &cluster(SI)      \\ [1ex]		    
 B458-D049      & 10407             & ACS/WFC  & F435W F606W                   &cluster  &cluster(SI)	   \\ [1ex]		    
 B475-V128      & 10818             & WFPC2    & F450W F814W                   &cluster  &cluster(SI)      \\ [1ex]		    
 B480-V127      &                   &          &                               &	 &cluster(SI) 	   \\ [1ex]		    
 B483-D085      &                   &          &                               &	 &cluster(SI) 	   \\ [1ex]		    
 B484-G310      &                   &          &                               &	 &cluster(SI)      \\ [1ex]		    
 B486-G316      &                   &          &                               &	 &cluster(S)  	   \\ [1ex]		    
B189D-G047      &                   &          &                               &	 &cluster(SI)      \\ [1ex]		    
VDB0-B195D      & 10818             & WFPC2    & F450W F814W                   &cluster  &cluster(SI)      \\ [1ex]		    
  NB21-AU5      & 10006             & ACS/WFC  & F435W                         &cluster  &cluster(SI)	   \\ [1ex]		    
      NB67      & 10818             & WFPC2    & F450W F814W                   &star     &star(SI)	   \\ [1ex]		    
      NB83      &  5907             &   WFPC2  & F555W F814W                   &star     &star(SI)	   \\ [1ex]		    
B006D-D036      &                   &          &                               &	 &cluster(SI) 	   \\ [1ex]		    
B012D-D039      &                   &          &                               &	 &cluster(SI)      \\ [1ex]		    
B015D-D041      & 10818             & WFPC2    & F450W F814W                   &cluster  &cluster(SI)      \\ [1ex]		    
B111D-D065      &  9794             & WFPC2    & F336W F439W F555W F675W F814W &cluster  &cluster(SI)	   \\ [1ex]   
B206D-D048      &                   &          &                               &	 &cluster(SI)	   \\ [1ex]   
B257D-D073      & 10818             & WFPC2    & F450W F814W                   &cluster  &cluster(I)       \\ [1ex]   
     DAO47      &                   &          &                               &	 &cluster(SI)	   \\ [1ex]   
      V031      & 10818(9709)       & WFPC2    & F450W F606W F814W             &cluster  &cluster(SI)      \\ [1ex]   

      
\end{longtable}


\scriptsize
\begin{longtable} {@{}lccccc@{}}
\caption{Classification of candidate young clusters listed in Tab.~2 of F05.}\\ 
\label{tab:app3}
\kill
\hline \hline
\\
Name  &   Obs-ID & Camera  & Filters  & Class HST & Class C09\\
\\      
\hline 
\endfirsthead
\hline
\endfoot
\\
B015-V204      &         &            &                           &	     &cluster(SI)	    \\ [1ex]	    
B030-G091      &  6671   & WFPC2      & F555W F814W               &cluster   &cluster(SI)	    \\ [1ex]	    
     B090      & 10260   & ACS/WFC    & F606W F814W               &cluster   &cluster(SI)	    \\ [1ex]	    
B101-G164      &         &            &                           &	     &cluster(SI)	    \\ [1ex]	    
B102           & 10260   & ACS/WFC    & F606W                     &star      &star(SI)\\ [1ex]	    
B117-G176      &  9087   & WFPC2      & F336W                     &cluster   &cluster(SI)	    \\ [1ex]	    
     B146      & 10118(5435)   & ACS/WFC, WFPC2    & F160BW F255W F300W F814W                     &cluster	&SLH           \\ [1ex]        
B154-G208      &  9087   & ACS/WFC    & F435W                     &cluster   &cluster(SI)	    \\ [1ex]	    
B164-V253      &         &            &                           &	     &cluster(SI)	    \\ [1ex]	    
B197-G247      &         &            &                           &	     &cluster(SI)	    \\ [1ex]	    
B214-G265      &         &            &                           &	     &cluster(SI)	    \\ [1ex]	    
B232-G286      &  8059   & WFPC2      & F300W F450W F606W F814W   &cluster   &cluster(SI)	    \\ [1ex]	    
B292-G010      & 10631   & ACS/WFC    & F435W F606W	          &cluster   &candidate	    \\ [1ex]	    
B311-G033      &  6671(11081)   & WFPC2      & F555W F606W F814W               &cluster	&cluster(SI)           \\ [1ex]        
B324-G051      &  6699   & WFPC2      & F555W F814W               &cluster   &cluster(SI)	    \\ [1ex]	    
B328-G054      &  6699   & WFPC2      & F555W F814W               &cluster   &cluster(SI)	    \\ [1ex]	    
B347-G154      & 10818   & WFPC2      & F450W F814W               &cluster   &cluster(S)	    \\ [1ex]	    
B423           & idate   &            &                           &	     &candidate \\ [1ex]	    
B468           &  5112   & WFPC2      & F555W F814W               &cluster   &cluster(I) 	    \\ [1ex]	    
NB16           & 10818   & WFPC2      & F450W F814W               &cluster   &cluster(SI)	    \\ [1ex]	    
B150D          &         &            &                           &	     &candidate \\ [1ex]	    
                                                        
\end{longtable}

\end{appendix}

\end{document}